\documentclass[12pt]{article}
\usepackage{graphicx}
\usepackage{amsmath,amssymb,amsfonts}
\usepackage{amsthm}
\usepackage{mathrsfs}
\usepackage{xcolor}
\usepackage{textcomp}
\usepackage{manyfoot}
\usepackage{enumerate}

\theoremstyle{definition}

\begin{document}
\begin{center}
\textbf{Damages and Materiality: Effects on voluntary disclosure}

\bigskip Miles B. Gietzmann\footnote{Bocconi University;
miles.gietzmann@unibocconi.it} and Adam J. Ostaszewski\footnote{London School
of Economics; a.j.ostaszewski@lse.ac.uk}

\bigskip
\end{center}

\textbf{Abstract. }    How should a court resolve a shareholder--management dispute following a materially significant price decline when it is suspected that management, at an earlier point in time, failed to update the market by disclosing a privately observed material event?
    A foundational result in this literature (Dye, 2017) shows that if a court publicly commits to increasing damages awards in an effort to deter nondisclosure, the policy may have a perverse effect: management may rationally choose to disclose even less. Schantl and Wagenhofer (2024) attribute this outcome to the pure insurance effect, whereby shareholders benefit from higher damages payments. They show that this result may be mitigated if management also face a fixed, exogenous reputational cost of nondisclosure. However, these reputational costs are independent of the model's equilibrium; furthermore, they assume that the court eventually observes the true state of the world with certainty (delayed omniscience by the court), and do not account for standards of materiality, which differ across legal systems.
    In contrast, we develop a dynamic continuous-time model in which both damages and materiality standards are endogenous. We show that, as damages awards increase, a previously unrecognized dynamic effect emerges: management rationally switch to a candid (full) disclosure strategy. Moreover, raising the materiality threshold induces this switch earlier, thereby increasing the extent of voluntary disclosure. Our analysis therefore demonstrates that regulators should recognize the complementary effects of damages and materiality standards. We further characterize what we term the legal consistency zone, in which higher damages awards, coupled with an appropriately chosen materiality standard, endogenously increase voluntary disclosure.

\bigskip

\noindent\textbf{JEL Classification}: K13, K22, G38, G39, D82, M41\newline%
\noindent\textbf{MSC Classification:} Primary 91G50, 91-10, 91B38, 91B44,
Secondary 91G80, 93E35, 60G25, 91B32, 91B99\newline\noindent\textbf{Keywords:
}Asset-price dynamics; voluntary aggregate-disclosure; litigation; scienter liability; legal
consistency; Markov piecewise-deterministic modelling.

\section{Introduction}

A fundamental principle of securities markets is that corporate disclosures should be timely and accurate. Securities regulations, such as Rule 10b-5 in the United States, allow investors to pursue claims for fraudulent misrepresentation when firms either disseminate misleading information or knowingly withhold price-relevant information, provided the applicable materiality standard is satisfied. This paper focuses on the latter case.

    Only a minority of securities litigation ultimately results in court-awarded damages (Mnookin, 2021); most disputes are resolved through negotiated settlements. Nevertheless, because such settlements are reached in the shadow of potential court rulings, the design of litigation mechanisms plays a central role in determining outcomes. Accordingly, we examine how legal institutions influence firms' voluntary disclosure incentives.

    Our analysis focuses on two interacting mechanisms that discourage management from withholding price-relevant information. First, in a dynamic setting, managers may have an incentive to establish a reputation for candid and timely disclosure by voluntarily revealing all privately observed signals, including adverse ones. Second, securities regulation provides shareholders with legal remedies when management delays disclosure of material adverse information. We model this regulatory environment through endogenous damages awards and materiality standards and study how these jointly shape firms' disclosure policies.

    Our analysis demonstrates that the leading static result of Dye (2017)---that an increase in damages awards may reduce voluntary disclosure---does not generally extend to a dynamic continuous-time setting. To capture the legal system requires a fundamental extension of the piecewise continuous deterministic framework of Gietzmann and Ostaszewski (2023), developed afresh in Section 3.2. We model the legal system by introducing various formats of both materiality standards and  damages. We show how these affect disclosure incentives when management receive private information at random times between successive mandatory reporting dates. During periods without voluntary disclosure, equilibrium firm value evolves according to an ODE (ordinary differential equation) derived in Section 3.2, which reflects the market's updating of beliefs in the absence of news. That is, if there continues to be no disclosure the firm value follows the ODE and provides management with a threshold value above which they disclose a received signal.

    Within this framework, under a sparing disclosure strategy (i.e. when management only disclose voluntarily their private signals above a dynamic cutoff), we recover Dye's static result. However, this result is not robust dynamically. As damages increase, firms may rationally switch from a sparing strategy to candid disclosure (all signals received are disclosed). Thus our principal contribution is to identify two aspects:
(i) that switching to candid disclosure behaviour sufficiently early cancels the Dye-paradox;
(ii) that early enough switching is brought about by joint choice of both materiality standards and damages.
Higher materiality thresholds induce this transition to candid earlier, thereby increasing voluntary disclosure. Damages and materiality standards are thus complementary rather than independent policy instruments.

    Our Main Theorem (Section 4) identifies what we term the legal consistency zone. This consists of two regions. The first is a Goldilocks interval, in which legal parameters are neither too high nor too low and firms optimally switch from sparing to candid disclosure, increasing voluntary disclosure. The second is an interval in which firms adopt candid disclosure immediately, without any switching. Both regions stand in contrast to Dye's disclosure-reduction result and demonstrate that stronger legal sanctions can increase, rather than decrease, voluntary disclosure in a dynamic environment.

    Figure 6 provides a graphical characterization of the legal consistency zone. The underlying condition is closely related to the valuation of finite-lived leased assets (or continuously compounded bond holdings). The analogy arises because, during periods of nondisclosure, firm value declines in a manner analogous to the economic depreciation of a finite-lived leased asset.

    Our continuous-time framework also allows us to examine more realistic specifications of damages awards. Existing disclosure models typically assume that courts eventually observe the private information withheld by management. We refer to this as the delayed omniscience assumption. Although analytically convenient, this assumption is difficult to justify in practice. Instead, we propose an alternative benchmark. Because our model characterizes the firm's equilibrium value immediately before the mandatory disclosure date when no voluntary disclosure has occurred, this observable equilibrium valuation can replace the unobservable delayed-omniscience benchmark when determining damages.

    The remainder of the paper is organized as follows. Section 2 reviews the related two-period disclosure literature. Section 3 develops the continuous-time model. Section 3.1 introduces the firm's information structure, while Section 3.2 derives the equilibrium valuation, the upper disclosure threshold, and the firm's liability for adverse valuation changes under litigation and settlement. Section 3.3 derives the lower disclosure threshold and characterizes the managerial trade-off between maintaining short-run market value and avoiding larger valuation declines at mandatory disclosure. Section 3.4 presents examples of explicit solutions to the equilibrium differential equation, and Section 3.5 examines disclosure probabilities and the emergence of switching between sparing and candid behaviour. Section 3.6 studies the effects of alternative materiality standards, while Section 3.7 considers time-varying liability shares that encourage earlier disclosure. Section 4 analyses litigation effects and establishes our Main Theorem on the legal consistency zone. Section 5 concludes. Appendix A contains the proofs and technical extensions, Appendix B summarizes the Brownian-motion results used throughout the paper, and Appendix C presents the detailed equilibrium formulation together with Mathematica code for simulating Poisson monitoring of firm value.

\section{Literature Review}

Dye in (Dye, 2017) studies a revised form of the standard two-period
disclosure problem of (Dye, 1985) in which a seller of an asset, who may
receive a private signal on the estimated intrinsic value of the asset,
decides whether to disclose the signal realization to potential buyers. The
revision (Dye, 2017) introduces is that after the sale a non-disclosing
seller is liable to damages specified by a damages-multiplier times the
amount the buyers of the asset overpaid for the asset. The overpayment is
calculated based on what the asset would have sold for, if the seller's
signal had been disclosed in the market. That is, the model of legal damages
assumes the seller-observed signal value is discovered by the courts after
the sales transaction is completed. Dye in (Dye, 2017) establishes several
counter-intuitive results within this simple model setting. He shows that
when the numerical value for the damages multiplier is increased this leads
to the seller being less likely to disclose the private signal, and this
decline in disclosure also follows if an independent fact-finder (auditor)
is more likely to detect non-disclosure. This leads Dye to conclude that his
result yields the important paradoxical conclusion that greater damages
payments for non-disclosure generate less disclosure (p. 706). The Dye result is sometimes described as
a pure-insurance effect because when investors are given litigation rights
to seek compensation, they incorporate its effect in the pricing of the seller
asset. Since investors rationally believe they are price protected against
non-disclosure, the investors become less cautious. As we shall see below,
our model makes a different prediction, because we show how raising damages
awards may incentivize a firm to switch earlier to a candid (full)
disclosure strategy away from a sparing (cutoff) strategy (see Main
Theorem). That is, although the cutoff rises for the sparing strategy (Proposition 4(b)) the
firm chooses to switch away from that strategy.

Schantl and Wagenhofer (2024) develop a Dye-type disclosure model that
formalizes how this lower disclosure cutoff can arise. They assume that
litigation, in addition to a pure insurance effect, also creates the
possibility of a \textit{deterrence effect}, in which the managers of firms
experience a reputation cost even though they have full litigation
insurance. This new deterrence effect introduces a dual cutoff strategy:
the traditional (upper) Dye cutoff and now a new lower cutoff below which
the reputation cost outweighs the valuation benefits from non-disclosure of
bad news. This dual cutoff strategy allows Schantl and Wagenhofer to show
that increasing legal damages payments for non-disclosure does not generate
less disclosure in the case that the reputation effect dominates the
insurance effect. However, they do not model how reputation arises from model parameters  in their setting. Another limitation is that, like (Dye, 2017), they assume that a signal observed privately by the firm's
manager, but not disclosed at the time of receipt, is observed perfectly by
the courts, but with subsequent delay. The difference between that value and
the market price of the firm, had the private signal been disclosed, then
determines the damages. That is, the courts are presumed to enjoy \textit{%
delayed but perfect observability}. A major result of Schantl and Wagenhofer
(2024) is that decreasing litigation risk leads to less disclosure of very
bad news, provided the reputation effect of non-disclosure is not too high.
In our model there is no exogenously imposed fixed reputation cost.

Choi and Spier (2022) introduce a dual (low, high) productivity type Dye
model in which public investment in low productivity firms is socially
inefficient. However, such investment can occur if firms privately observe
low project productivity but choose during fund-raising to not disclose it
(\textquotedblleft withholding material information\textquotedblright) and
pool with uninformed firms. At issue is how to design the legal damages
system which specifies how firms are liable for non-disclosure to deter
inefficient investment. Choi and Spier show how in equilibrium firms'
non-disclosure depends on the frequency with which the firms are privately
informed and the level of liability. They show that as the liability system
becomes stronger, low productivity firms are increasingly deterred from
raising funds. In the case where damages for non-disclosure only offer
partial deterrence to strategic firms, they show how social welfare
increases in damages. Like (Dye, 2017) they apply a delayed omniscience
assumption for the courts, by modelling damages as being assessed on the
realized value of the private productivity signal at the time of the court
hearing (not disclosed by the firm at the (earlier) time of observation).
They also look at three extensions: firm managers being personally liable,
induced false positives, and costly litigation.

\section{Model framework}

We exploit here the earlier modelling in Gietzmann and Ostaszewski (2023),
but now amended so as to include possible shareholder litigation,
re-deriving a continuous-time framework in which again during periods of
silence an ODE (ordinary differential equation) describes the equilibrium
evolution of the \textit{market value} (Section 3.2) and the detailed structure of the ODE responds to changes in the legal system defined by various regulatory combinations of materiality standards and of damages. This evolution is
associated with equilibrium market inferences, such inferences considering
whether falls observed in price go beyond trend values, thereby suggesting
the possibility that the firm withheld negative private information.

All values referred to below are taken to be \textit{discounted} to the
starting date $t=0,$ the initial mandatory disclosure date, and the
reporting period ends with a mandatory terminal disclosure date of $t=1$.

\subsection{Observations and Damages threshold}

An internally generated signal of the firm's value, viewed as the basis for
forecasts of future value, denoted $Y_{t}\text{,}$ is privately observed
intermittently between the mandatory disclosure dates so that the \textit{%
random} observation times $t$ of $Y_{t}$ follow a Poisson process of
intensity\textbf{\ }$\lambda ,$ independent of the signal value and not
known to the market. We model the signal, consistently with the benchmark
Black-Scholes framework, as a geometric Brownian \textit{martingale}$\text{,}
$ so that its log-value is described by
\begin{equation*}
\log Y_{t}=\log Y_{0}-\frac{1}{2}\sigma ^{2}t+\sigma W_{t}\text{,}
\end{equation*}%
where $W_{t}$ is standard Brownian motion under a Borel measure $\mathbb{Q}$
on $\mathbb{R}.$ Here the volatility factor $\sigma $ aggregates volatility
in the economic production process and in the precision of its evaluation.
We denote by $\{\mathcal{Y}_{t}\}_t$ the management's private information
filtration accumulating over time $t$. (For the explicit definition of the
private filtration see Appendix B.)

If observed and if disclosed at time $t$, the signal $Y_{t}$ will be
disclosed truthfully and so establishes a new valuation of the firm at that
point in time as $Y_{t}$, there being no other source of information about
the firm. For this reason we term $\mathbb{Q}$ the \textit{market measure}.
However, absent any disclosure, the market will remain uninformed as to
whether a signal has been received. The public (market) information filtration
accumulating over time $t$ is denoted by $\{\mathcal{F}_{t}\}_t$. (See Appendix B
for the formal relation of the public to the private filtration, when
management seeks to maintain the firm's valuation as high as possible;
management are assumed to have interests aligned with those of the
shareholders.)

The ambiguity of silence (over a lack of signal or of a non-disclosed
signal) at any point in time may be viewed as a continuous analogue of the
\textit{scheduled} `potential' interim announcement in the foundational
framework of (Dye, 1985). On any interval of such silence, this ambiguity
yields a valuation $\gamma _{t}$\ of the firm at each time $t,$\ on the
assumption that management pursue maximal valuation on grounds of the same
equilibrium considerations as in (Dye, 1985); it will be characterized below
(Section 3.3) as the solution of an ODE and represents a threshold for
disclosing at time $t<1$\ an observation of $Y_{t}$ above\textbf{\ }$\gamma
_{t}.$\textbf{\ }

The ODE has a explicit solution (in terms of the model's parameters, such as
the arrival rate $\lambda $ of private news and the combined observation and
production variability $\sigma $). Throughout a period of silence it
expresses in implicit terms (absent any voluntary disclosures) an
equilibration of two opposing forces: non-disclosure of what might be
interpreted as bad news, protecting the value (before the terminal time of $%
t=1$) of the firm, versus a higher likelihood that litigation will require
damages to be paid.

Its solution will be an exponential-like declining continuous convex curve $%
\gamma _{t}$ recalibrated so that the last known disclosed value, assumed to
occur at time $t=0,$ has $\gamma _{0}=1.$ As $t$ increases the associated
\textit{equilibrium trend curve} $\gamma _{t}$ is the notional principal component in
the market's assessment of firm value when there is no (corrective)
disclosure. This reflects the market's inference from continued silence as
an increased likelihood (but not certainty) of an undisclosed poor signal
value having been privately received by the firm. The date of the next
mandatory disclosure is taken to be $t=1.$ The \textit{limiting value} as $t$
tends to $1$ of $\gamma _{t}\text{,}$ namely%
\begin{equation*}
\gamma _{1}=\lim_{t\rightarrow 1^{-}}\gamma _{t}\text{,}
\end{equation*}%
now has an important significance: it is the market's valuation just prior
to the mandatory disclosure itself, the later mandatorily disclosed
valuation being $Y_{1}$. Any fall in price that might occur from $%
\gamma _{0}$ to $Y_{1}$ is decomposed into two components - first an \textit{%
expected} fall under continued silence $\gamma _{0}-\gamma _{1}$ and an
\textit{unexpected} component $\gamma _{1}-Y_{1}$. The innovation in our
approach is to model allowed litigated claim levels as being set to depend
only on the unexpected component, namely the difference between an equilibrium
value (under silence) inferred (calculated) by the market and an observed
realized value $Y_{1}$. Thus $\gamma _{1}$ becomes the \textit{damages
threshold.}

\subsection{Equilibrium and the upper cutoff}

We can now derive the determining equation for $\gamma _{t}.$ Below, though
we work in a continuous framework, we follow the (Dye, 1985) paradigm\footnote{Our approach
here follows the disclosure literature in deriving equilibrium behaviour from determining a disclosure cutoff defined by indifference between disclosure and non-disclosure. A brief description follows in this subsection
and a more detailed formulation is in Appendix B.} of
three dates (in his case the mandatory initial-, the voluntary interim-, and
the mandatory terminal-disclosure dates) from which to derive a downgrade at
the interim date. Relocating these three dates respectively to $t<s<1\text{,}$ the
interim downgrade $\gamma _{s}$, which is also the threshold for
announcements in equilibrium at time $s\text{,}$ is the value $\gamma
=\gamma _{s}$ which symbolically satisfies
\begin{align*}
\gamma =\mathbb{E}^{\mathbb{Q}}(Y_{s}|ND_{s}(\gamma ),\mathcal{F}_{t}),
\tag{indif}
\end{align*}%
where as above $\mathbb{Q}$ is the market measure and $\mathcal{F}_{t}$ is
the market information accumulated by time $t$;\newline
$ND_{s}(\gamma )$ is the event at time $s$ that no disclosure occurs when
the signal is below $\gamma $;\newline
$RHS=$ the market's expectation of value, conditional on the non-disclosure
event $ND_{s}(\gamma _{s})\text{,}$ including expected damages-settlement.

We assume here that the observation $Y_{s}$ is of the financial state of the
firm, treated as a hidden variable (see Gietzmann and Ostaszewski (2016)
section 2.3), and that the regression function $m_{s}(y):=\mathbb{E}_{t=0}^{%
\mathbb{Q}}[Y_{1}|Y_{s}=y]$ is increasing. Then the optimal threshold $%
\gamma _{s}$ is uniquely determined and has three properties as follows.
\medskip

(i) \textbf{Minimum Principle }(Ostaszewski and Gietzmann (2008), cf.
Acharya et al. (2011)): The valuation function
\begin{equation*}
W(\gamma ):=\mathbb{E}_{0}^{\mathbb{Q}}[Y_{1}|ND_{s}(\gamma )]
\end{equation*}%
has a unique \textit{minimum} at $\gamma =\gamma _{s}$. This result implies
the Minimum Principle: \textit{In equilibrium the market values the firm at
the least level consistent with the beliefs and information available to the
market as to the firm's productive capability.}

See Gietzmann and Ostaszewski (2023) for this form of the so-called \textit{%
Unravelling Principle} due originally to Grossman-Hart-Milgrom.
\medskip

(ii) \textbf{Risk-neutral Consistency Property: }$\gamma _{s}$ is the unique
value $\gamma $ such that%
\begin{equation*}
\mathbb{E}_{0}^{\mathbb{Q}}[Y_{1}|Y_{0}]=\tau _{\text{D}}\mathbb{E}_{0}^{%
\mathbb{Q}}[Y_{1}|\text{disclose }Y_{s}\geq \gamma ]+\tilde{\tau}_{\text{D}%
}m_s(\gamma ),
\end{equation*}%
where $\tau _{\text{D}}$ is the (time $t=0$) market probability of
disclosure occuring at $s$ and $\tilde{\tau}_{\text{D}}=1-\tau_{\text{D}}$. This is highly significant, in that the
valuation at time $0$ anticipates the potential effects of a voluntary
disclosure at the future interim date $s$. In brief, the approach is
consistent with the principles of \textit{risk-neutral valuation}; for
background see (Bingham and Kiesel (1998), Chap. 6).
\medskip

(iii) \textbf{Interim downgrade: }From the perspective of time $t,$ in a
model with only three dates: $t<s<1,$ the Dye equation at time $s$ may be
written:

\begin{align*}
\gamma _{s}=
\frac{\tilde{q}\gamma _{t}+q\int\nolimits_{0}^{\gamma _{s}}yd%
\mathbb{Q}_{t}(y)+q\int\nolimits_{0}^{\gamma _{s}}\gamma _{s}R_{s}(1-y/\gamma
_{s})d\mathbb{Q}_{t}(y)}{\tilde{q}+q\mathbb{Q}_{t}(\gamma _{s})},  \tag{cond}
\end{align*}%
as $\gamma _{t}=\mathbb{E}_{t}^{\mathbb{Q}}[Y_{1}]\text{,}$ where the $t$-subscript
indicates conditioning on $\mathcal{F}_{t}\text{.}$ Here the probability of
a signal being observed by management at time $s$ is $q=q_{s}\text{,}$
assumed exogenous and independent of the state of the firm, and $\tilde{q}%
:=1-q.$ The third term in the numerator recognizes the damages received by
shareholders on the grounds that some signal $y<\gamma _{s}$ is likely to
have been observed at the time $s$ or earlier and disclosed; this is
expressed here by a general proportionate \textit{restitutionary}
\textit{damages schedule} $R_{s}(x),$ assumed non-negative, continuous in $s$
and smooth in $x$, applied to
\[
x=(\gamma -y)/\gamma =1-y/\gamma >0,
\]%
i.e. to the proportionate valuation fall relative to the expected value $%
\gamma $ when the disclosed signal has value $y<\gamma .$ Our analysis will
presently lead to various natural choices of a function $R_{s}(x),$ when
passing to the `instantaneous' version of (cond), i.e. an ODE, obtained in the limit as $s
$ approaches $t.$ One such choice is a fractional damages-multiplier, as
proposed in (Dye, 2017), namely  $R_{s}(x)\equiv \rho _{s}x,$ with $0\leq
\rho _{s}\leq 1,$ identifying damages as a liability share. The factor $\rho
_{s}$ is independent of the signal of time $s$ (i.e. assuming $Y_{s}$ is
observed) and may take into account characteristics of the firm as well as
the time $s$ (i.e. promptness of a disclosure).
\bigskip

We may now adopt the received standard approach in determining a perfect
Bayesian equilibrium as in the expositions of (Dye, 2017) and (Dye and
Sridhar, 2024) in a backwards-dynamic programming fashion. We cite their
statements nearly verbatim. As there, our calculations are based on
risk-neutral evaluations (so the securities markets are competitive in the
sense that the price of the firm is a martingale and equals the expected
value of the firm's future economic value at all points in time based on the
publicly available information at that time); the manager does not
independently participate in the securities markets; the investors comprise
representative agents (are homogeneously informed) and obtain all their
information from the firm's filings.

Thus prices are set correctly based on the firm's disclosures or
non-disclosure, and the firm's disclosure decision is optimal, taking the
prices the firm believes the representative investors will pay for the asset,
conditional on disclosure or non-disclosure as given. Here too we make the
Grossman-Milgrom assumption that the firm can never make a false disclosure,
the specification of prices given in equation (cond) applies both
\textquotedblleft on\textquotedblright\ and \textquotedblleft
off\textquotedblright\ the equilibrium path, that is, whether or not buyers
expected the firm to disclose its observed signal.

We now state what should be regarded as a limiting restatement in continuous time of the Dye indifference equation.

\bigskip
\textbf{Proposition 1 (Equilibrium value dynamics).} \textit{The
continuous-time limit of the Dye equation }(indif)\textit{\ incorporating
Poisson-directed private observations and an assumed smooth damages schedule $R_{t}(.)$ takes the general form}%
\begin{align*}
\gamma_{t}^{\prime}=-\lambda(1-\rho_{t}^{*})\gamma_{t}h(t),  \tag{cont-eq*}
\end{align*}
\textit{for some }$\rho^{*}_{t}:=R^{\prime}_{t}(1-y^{*}_{t}/\gamma_{t})$
\textit{ with }
$0\leq y^{*}_{t}\leq \gamma_{t},$ \textit{where}
\[
\gamma _{t}h(t)=\int_{0}^{\gamma _{t}}(\gamma _{t}-y)d\mathbb{Q}_{t}(y\mathbb{%
)}=\gamma _{t}\int_{0}^1\mathbb{Q}_{t}(\gamma _{t}y)dy,
\]
\textit{after integration by parts and rescaling. When }$R_{t}(x)\equiv \rho _{t}x,$\textit{\ with }
$0\leq \rho_{t}\leq 1,$\textit{\ the differential equation reduces to}
\begin{align*}
\gamma_{t}^{\prime}=-\lambda\gamma_{t}(1-\rho_{t})h(t).  \tag{cont-eq}
\end{align*}%

More generally, $h(t)$ is the instantaneous value of a protective put against valuation falls\footnote{Above $h$ is the `lower first partial moment below the target $\gamma_{t}$', briefly the \textit{hemi-mean.}}, as determined by the legal system in force, under sparing equilibrium behaviour at time $t$ and otherwise 0 under candid. An endearing feature of the second integral characterizing $h(t)$ is that the probability $\mathbb{Q}_{t}$ can recognize a different support as determined by the legal system (for example\footnote{%
This depends on the legal framework dictating the form of $R_{t}(.).$ See
subsection 3.6 for an alternate framework.}, under a materiality standard, assigning 0 probability to non-material claims).

 Here\ $\rho _{t}$ is any continuous-time liability-share factor, as studied in detail in
Section 3.6. Under sparing behaviour throughout and with all claims below the damages threshold actionable, the function $h$ is the \textit{instantaneous decline} described by
\begin{equation*}
h_{\text{vanilla}}(t)=2\Phi (\sigma /2\sqrt{1-t})-1.
\end{equation*}%

\noindent As earlier, $\sigma $ represents the \textit{(aggregate) volatility}
(aggregating productive and observation vols.). For a proof (absent any
damages) see Gietzmann and Ostaszewski (2016) and Gietzmann and Ostaszewski
(2023). The formula above for $h(t)$ evaluates the limiting protective put at
time $t$ (expiring at time  $t=1$), using the standard  Black-Scholes setting:
\begin{align*}
P_{\gamma _{s}}=C_{\gamma _{s}}+\gamma _{s}-\gamma _{t} \rightarrow \gamma
_{t}\Phi (d_{+})-\gamma _{s}\Phi (d_{-})\rightarrow \gamma _{t}h_{\text{vanilla}}(t),
\end{align*}%
and decreases with time. For us this is a benchmark case of protective put against a fall in price, and so denoted and referred to as the (B-S) \textit{ vanilla } put. Indeed, reference to the
Poisson probability of a jump occurring in $[t,s]$ as being $%
q=q_{ts}=\lambda (s-t)+o(s-t),$ using the Landau little-oh notation, and
with $\tilde{q}=1-q_{ts},$ after some re-arrangements (involving integration by
parts -- Appendix A), equation (cond) leads to:
\begin{align*}
(1-q_{ts})\frac{\gamma _{s}-\gamma _{t}}{(s-t)}&=-\lambda \int\nolimits_{y\leq \gamma _{s}}\mathbb{Q}_{t}(y)\text{{}}%
\mathrm{d}y\text{,}\\
+&\lambda R^{\prime}_{s}(1-y^{*}_{s}/\gamma_{s})\int\nolimits_{y\leq \gamma _{s}}\mathbb{Q}_{t}(y))\text{{}}%
\mathrm{d}y\text{,}
\end{align*}%
by the mean-value theorem for integrals (applicable as $R^{\prime}_{s}(.)$ is assumed continuous and $\mathbb{Q}_{t}$ does not change sign), for some $y^{*}_{s}\in (0,\gamma_{s})$, or with $\rho^{*}_{s}:=R^{\prime}_{s}(1-y^{*}_{s}/\gamma_{s})$
\begin{equation*}
(1-q_{ts})\frac{\gamma _{s}-\gamma _{t}}{(s-t)}=-\lambda (1-\rho
_{s}^{*})\int\nolimits_{y\leq \gamma _{s}}\mathbb{Q}_{t}(y)\text{{}}%
\mathrm{d}y\text{,}
\end{equation*}%
ignoring errors of order $o(s-t)/(s-t)\text{.}$ Passing to the limit as $%
s\downarrow t$ we obtain the limiting formulas above.

Armed with these definitions, an equilibrium comprises for each time $t:$
prices that are either $Y_{t}$ if the signal is received and disclosed but
otherwise $\gamma_{t}$, which is a time $t$ cutoff for disclosure of
observations of $Y_{t}$, and a \textquotedblleft no
disclosure\textquotedblright\ set $ND_{t}$ such that (cf. (Dye and Sridhar, 2024)):\newline
\noindent(i) if the firm discloses $Y_{t}$, then the investors price the
stock at $Y_{t}$;\newline
\noindent(ii) given that investors think the firm uses the cutoff $%
\gamma_{t} $, if the firm makes no disclosure, then investors price the
asset at $\gamma_{t}$;\newline
\noindent(iii) given that the firm thinks investors set the price of the
stock when the firm discloses $Y_{t}$ as being $Y_{t}$ and when the firm
makes no disclosure as being $\gamma_{t}$, the firm discloses the estimate $%
Y_{t}$ when it receives the observation if and only if $Y_{t}\geq\gamma_{t}$%
, where $\gamma_{t}$ is as defined by (cond), and the set $ND_{t}$ is given
by $ND_{t}=\{Y|Y>\gamma_{t}\}$.

In the light of this analysis, sparing behaviour consists in disclosing $%
Y_{t}$ if observed at time $t$ provided $Y_{t}>\gamma _{t}.$ (Appendix B
offers a more formal account of the equilibrium considerations.) Figure 1
summarizes the various cutoffs.
\begin{figure}[tbp]
\begin{center}
\includegraphics[height=3.5cm]{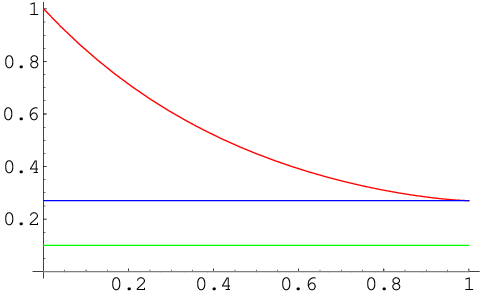}
\par
{Figure 1. Disclosure cutoffs (p.v.) over time interval (0,1): red upper
(equilibrium trend value $\gamma _{t}$), blue lower (precaution against
damages liability for threshold $\gamma _{1}$) and green lowest (precaution
against managerial personal liability)}
\end{center}
\end{figure}

\subsection{Explicit solutions and simple examples}

In the case of a firm silent until the mandatory disclosure date, i.e.
silent in the interval is $[0,1),$ we may solve%
\begin{equation*}
\gamma _{t}^{\prime }=-\lambda \gamma _{t}h_{\rho }(t)\text{ with }h_{\rho
}(t):=(1-\rho _{t})h(t),
\end{equation*}%
by separation of variables: take $\displaystyle{g_{\rho }(t)=\int_{0}^{t}h_{\rho }(u)\text{ }\mathrm{%
d}u}$, then
\begin{equation*}
\log \gamma _{t}=\int_{0}^{t}\frac{d\gamma _{t}}{\gamma _{t}}=-\lambda
g_{\rho }(t):\qquad \gamma _{t}=e^{-\lambda g_{\rho }(t)}.
\end{equation*}%
In particular, this yields the left limit $\gamma _{1}$ to be higher than
that obtained when no litigation is permitted:
\begin{equation*}
\gamma _{1}=\lim_{t\rightarrow 1}\gamma _{t}=e^{-\lambda g_{\rho
}(1)}>e^{-\lambda g(1)}:=\gamma _{1}^{\text{no lit}}.
\end{equation*}%
More generally (Proposition 4(a) and 4(b) below), the whole of the upper disclosure curve arising under
litigation lies above the corresponding curve when litigation is prohibited,
for which $\gamma _{t}^{\text{no lit}}=e^{-\lambda g(t)}$.

The case $\rho _{t}\equiv \rho ,$ a constant, is particularly transparent:
\bigskip

\noindent\textbf{Proposition 2 (Constant liability-share: comparative
statics). }\textit{Under the sparing regime the valuation curve }$\gamma_{t}$%
\textit{\ is higher for a higher constant factor }$\rho$.\textit{\ In
particular, the damages cutoff is higher for a constant higher factor }$\rho
$\textit{.}
\medskip

Indeed, if $\rho_t^H\equiv\rho^{H}>\rho^{L}\equiv\rho_t^L,$ then $\gamma_t^H:=e^{-\lambda(1-\rho^{H})g(t)}>e^{-%
\lambda(1-\rho^{L})g(t)}=\gamma_t^L.$

\subsection{A lower cutoff and a conflict of interest}

In this section we identify how the \textit{damages threshold} $\gamma _{1}$
provides a cutoff below which observed signals represent what we term `bad news'. This lower cutoff on the signal range builds a
fair-pricing relationship between such bad news and the upper cutoff of the preceding
section and identifies a conflict of interest between maintaining high value
and avoiding high damages.

In Proposition 3 below we refer to the liability-share formula for damages: $R_{t}(x)=\rho_t x$ and the notion of equilibrium valuation of Proposition 1. We recall that this equilibrium is based on Dye's valuation indifference.

We will presently see that management may prefer to avoid dispute settlements by
voluntarily disclosing any earlier observation $Y_{t}<\gamma _{1}\text{.}$
In view of this we will also study the effects of sensitivity to early disclosure,
whereby $\rho $ may depend on the timing $t$ (how early) the disclosure
is made and so replaced by a \textit{timed version} liability-share $\rho
_{t}$, and lastly we will compute the probability of observations being not
disclosed, given the structure of the firm in regard to pay-for-performance
and intelligence gathering.

\bigskip
\noindent \textbf{Proposition 3 (Effect of equilibrium damages on valuation}%
; cf. (Schantl and Wagenhofer (2024)). \textit{Assume the court threshold
for awarding damages is }$\gamma _{1},$\textit{\ appropriate personal
liability is borne by management and that damages at time t=1 are awarded according to a liability-share formula:}
$$\rho_1(\gamma_1-Y_{1}).$$
\textit{In the equilibrium of Proposition 1, if at any time }$t<1$
\textit{the observed signal is }$Y_{t}<\gamma _{1},$ \textit{then at that
time }$t$ \textit{the management's (net-of-damages) private valuation, i.e. conditioning on $\mathcal{Y}_{t}$, falls below }$Y_t $
\textit{to}
\begin{equation*}
(1+\rho )Y_{t}-\rho (\gamma _{1}+\mathbb{E}_{t}^{\mathbb{Q}}[(Y_{1}-\gamma
_{1})^{+}|Y_{t},\mathcal{Y}_{t}]).
\end{equation*}%
\textit{Thus disclosure of }$Y_{t}$\textit{\ leads to a fall in market valuation below the equilibrium upper limit }$\gamma
_{t}.$
\bigskip

\noindent\textbf{Corollary 1. }\textit{The corresponding lower cutoff is }$%
\delta_{t}\equiv\gamma_{1}\text{.}$

For the proof see Appendix A. The management of the firm thus faces a \textit{%
conflict of interest} between maintaining a higher market valuation through
sparing rather than candid behaviour versus a higher damages settlement;
this is resolved by an appropriate management objective function, treating $%
\gamma _{t}$ as a state variable controlled by the optimal probabilistic
managerial disclosure choice. This, like the sparing disclosure behaviour,
comes from assumed alignment of managerial with shareholder interests, in
line with the assumptions in (Dye, 1985). We delay considering this
trade-off till Section 4 to give precedence to a study of the
liability-share factor as that is intimately linked with the governing
equation for $\gamma _{t},$ each potentially influencing the other.
Proposition 1 stands in contrast to the prospective effects in the so-called
US Reform Law of 1995: see Dybvig et al. (2000).

\subsection{The `no-litigation' example $\protect\rho \equiv 0$}
A special case which provides initial insights and is a building block
towards more general behaviours corresponds to $\rho \equiv 0$ (no damages and so no
litigation). This is the simplest context which
answers the natural question of what is the probability of
suspected non-disclosure, as the probability depends only on $\gamma _{1}$
and not on the expected damage-losses.
Here there exist two equilibria involving a single switch between the two
kinds of behaviour: sparing and candid, both observed empirically cf.
(Grubb, 2011). This plays on a firm having already established a reputation
(recognized by the market) for selection of these switching behaviours in
equilibrium. The context most conveniently starts with that of a
sparing-first behaviour where, after an optimal switching time $\theta \text{%
,}$ management switch to being candid thereafter; if no switching is
involved, one may take $\theta =0$ to create a context of sparing-only
behaviour. Subsequently, we consider a switch at an optimal time $\theta $
from candid-first behaviour to sparing behaviour. We defer the calculations supporting the formulas
used here to Section 4. The key point is that we shift origin and rescale
via the transformation
$$s_{\theta}(t)=(t-\theta)/(1-\theta),$$
which allows the interval $[\theta,1]$ to reproduce
sparing behaviour as modelled on $[0,1]$. See Proposition 5 in Section 4 for the general `with
litigation' characterizations. Here in the candid-first context we take $h(t)=0$ for $t\in [0,\theta)$ (candid disallows non-disclosure protection) and
$h(t)=h_{\theta}(t)$ for $t\in [\theta,1]$, where
\[
h_{\theta }(t):=h_{\text{vanilla}}((t-\theta )/(1-\theta )).
\]
The solution of (cont-eq) in, respectively, the sparing-first
and candid-first modes is as follows:%
\begin{equation*}
\gamma _{t}=\left\{
\begin{array}{cc}
e^{-\lambda g(t)}, & 0\leq t\leq \theta  \\
\gamma _{\theta }=e^{-\lambda g(\theta)}, & \theta \leq t<1\text{;}%
\end{array}%
\right.
\end{equation*}%
\begin{equation*}
\gamma _{t}=\left\{
\begin{array}{cc}
1, & 0\leq t\leq \theta , \\
e^{-\lambda g(s_{\theta}(t))]}, & \theta \leq t<1.%
\end{array}%
\right.
\end{equation*}%
The latter case satisfies $\gamma _{t}^{\prime }=-\lambda h_{\theta}(t)\gamma _{t}%
\text{ in the second subinterval. Here}$
\begin{equation*}
\gamma _{1}=e^{-\lambda g(1)}.
\end{equation*}%
With $T=1-\theta $ or $T=\theta $ being the length of the sparing interval,
according to context, the probability of a Poisson arrival in the interval
is $1-e^{-\lambda T}$.

Consider the sparing-first strategy. The optimal switching point $\theta $ solves
the equation%
\begin{equation*}
(1-\theta )h(\theta )=(\kappa ^{-1}-1)/\lambda
\end{equation*}%
where $0<\kappa<1$ is a managerial performance parameter (refer to the no-litigation study, or the more general Proposition 5 below).
So, as $(1-t)h(t)$ is decreasing in $t\text{,}$
\begin{equation*}
\lambda =\frac{(\kappa ^{-1}-1)}{(1-\theta )h(\theta )}\geq \frac{\kappa
^{-1}-1}{h(0)}=: \lambda _{\min }^{\mathrm{spar}}.%
\end{equation*}%
Here $\theta =0$ for $\lambda =(\kappa ^{-1}-1)/h(0)$, so that low values of
$\lambda $ correspond to low $\theta $ and so candour almost throughout.
Large values of $\lambda $ increase the sparing duration $\theta $, leading
to a lowering of the terminal threshold $\gamma _{1}$ which tends to $0$ as $%
\lambda \rightarrow \infty .$

Since we model $Y_{t}$ as a martingale and $Y_{0}=1\text{,}$ $\log
Y_{t}$ for $\theta <t<1$ is represented by%
\begin{equation*}
\log Y_{t}=-\frac{1}{2}\sigma ^{2}t+\sigma W_{t}\text{,}
\end{equation*}%
with $W_{t}$ a standard Brownian motion. If no observations in $(0,1)$ are
reported, then any withheld observations occurred in $(0,\theta ]\text{.}$
Recalling that $\gamma _{\theta }=e^{-\lambda g(\theta )}\text{,}$ the
probability that $Y_{t}<\gamma _{\theta }$ for some Poisson arrival time $t$
in $[0,\theta ]$ is given by the probability that for some Poisson arrival
time $t$ with $0<t\leq \theta $, when (on dividing by $\sigma$) $\displaystyle{-\dfrac{1}{2}\sigma t+W_{t} <-\lambda g(\theta )/\sigma}$, i.e.
\begin{equation*}
\min_{\tau \in \lbrack 0,\theta ]}\left( -\dfrac{1}{2}\sigma \tau
+W_{\tau }\right) <-\lambda g(\theta )/\sigma ,
\end{equation*}%
where $\tau $ is allowed only to be a Poisson arrival time. As before,
we use $\gamma _{1}=\lim_{t\rightarrow 1}\gamma _{t}$, with $\gamma _{t}$
the upper disclosure threshold. The probability of the event at time $\tau $%
, corresponding to a signal observation below $\gamma _{1},$ must then be
multiplied by the probability of a Poisson arrival time occurring, which is $%
1-e^{-\lambda \theta }\text{.}$

As there is no simple closed formula for this overall probability of the
event that management withhold privately observed signals, we have included
in Appendix B a simple Mathematica code which enables us to include here by
way of illustration how that probability depends on the intensity $\lambda$
for fixed $\sigma$ and $\kappa\text{.}$ Ee have taken $\sigma=3$ and $%
\kappa=0.7\text{.}$ The graph in Fig. 2a is obtained by using a Bezier spline through a
list of calculated points.

We have computed with 100 points each at distance apart of 1/10 to
obtain minimal `bending'. Note that the probability of non-disclosure of a
signal observed below $\gamma _{1}$ tends to $0$ as $\lambda \rightarrow
\infty $, since $\gamma _{1}$ tends to $0$ as $\lambda \rightarrow \infty .$
We may comment also that in equilibrium for $\lambda $ large the investor
will not believe that the manager is uninformed!

\begin{figure}[tbp]
\begin{center}
\includegraphics[height=3.5cm]{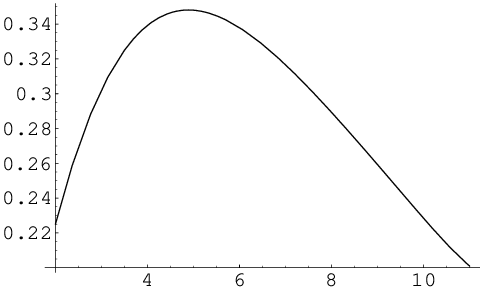}
\par
{Figure 2a. Scienter probability (withholding earlier signals below $%
\gamma_1 $) in Poisson sampled sparing-first case with $\sigma=3,\kappa=0.7.$%
}
\end{center}
\end{figure}

For comparison purposes consider the probability that%
\begin{equation*}
\min_{t\in \lbrack 0,\theta ]}\left( -\frac{1}{2}\sigma t+W_{t}\right)
<-\lambda g(\theta )/\sigma \text{,}
\end{equation*}%
where the times $t$ are not restricted to be Poisson arrival times. In this
case a closed formula is well-known (see Appendix A). In the formula we take
$a=-\lambda g(\theta )/\sigma <0$ and $\mu =-\frac{1}{2}\sigma \text{.}$ Now
the required probability is%
\begin{equation*}
\Phi \left( \frac{-\lambda g(\theta )+\frac{1}{2}\sigma ^{2}\theta }{\sigma
\sqrt{\theta }}\right) +e^{\lambda g(\theta )}\Phi \left( \frac{-\lambda
g(\theta )-\frac{1}{2}\sigma ^{2}\theta }{\sigma \sqrt{\theta }}\right)
\end{equation*}%
with $(1-\theta )h(\theta )=(\kappa ^{-1}-1)/\lambda$. Multiplying this by $1-e^{-\lambda \theta }$ yields the probability of an
observation below $\gamma _{\theta }\text{.}$ This is graphed in Fig. 2b for a
fixed $\kappa $ against $\lambda \geq \lambda _{\min }^{\mathrm{spar}}\text{.%
}$

\begin{figure}[tbp]
\begin{center}
\includegraphics[height=3.5cm]{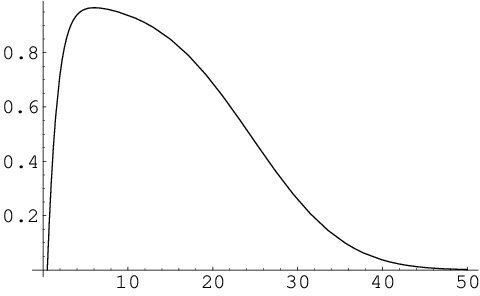}
\par
{Figure 2b. Same witholding probability for continuously monitored
sparing-first case with $\sigma=3,\kappa=0.7$}
\end{center}
\end{figure}

In this case increasing $\kappa$ increases the maximal probability of
withheld observation (as will be seen later, $\kappa=1-\alpha/\beta\text{,}$ so increasing $\beta$
for $\alpha$ fixed, i.e. stronger penalization of over-zealous candour
yields higher probability of withheld observations); increasing $\sigma$
likewise increases this maximal probability.

Small $\lambda$ yields lengthier candour. Large values of $\lambda$ yield a
lowering of the terminal threshold, explaining the low probability of
withheld observations.

In a \textit{candid-first strategy } the optimal switching time $\theta \leq
1$ is given by%
\begin{equation*}
g(\theta )=(\log \kappa ^{-1})/\lambda \text{.}
\end{equation*}%
(Refer again to the no-litigation study, or Proposition 5 below.) AS $%
g $ is a convex increasing function, $\theta $ is defined for $\lambda
\geq \lambda _{\min }^{\mathrm{cand}}=-\log \kappa /g(1);$ hence $\theta =1$
for $\lambda =-\log \kappa /g(1)$. So low values of $\lambda $ correspond to
candour almost throughout. Large value of $\lambda $ reduce $\theta $
towards $0\text{,}$ thus reducing the duration of candour and lead to almost
sparing behaviour throughout and also a lower terminal cutoff $\gamma _{1}$,
tending to $0$ as $\lambda \rightarrow \infty .$

Recalling again that we model $Y_{t}$ as a martingale with $Y_{0}=1\text{,}$
so that $\log Y_{t}$ for $\theta<t<1$ is represented by%
\begin{equation*}
\log Y_{t}=-\frac{1}{2}\sigma^{2}t+\sigma W_{t}
\end{equation*}
for $W_{t}$ a standard Brownian motion. In a period of silence $[0,1)$ in
which candid behaviour till time $\theta$ has brought no disclosure, to
compute the probability that $Y_{t}<\gamma_{1}$ for some Poisson arrival
time $t$ in $[\theta,1]$ we change origin and use time $s=t-\theta>0.$
Recall also that $\gamma_{1}=e^{-\lambda(g(1)-g(\theta))}\text{.}$ Thus the
required probability is that for some Poisson arrival time $s$ with $0\leq
s\leq1-\theta$ and $\displaystyle{-\dfrac{1}{2}\sigma s+W_{s} <-\lambda\lbrack g(1)-g(\theta)]/\sigma,}$ i.e.
\begin{equation*}
\min_{\tau\in\lbrack0,1-\theta]}\left( -\dfrac{1}{2}\sigma\tau+W_{\tau
}\right)  <-\lambda\lbrack g(1)-g(\theta)]/\sigma\text{,}
\end{equation*}
where $\tau$ is restricted to be a Poisson arrival time. The probability of
this event must then be multiplied by the probability of a Poisson arrival
time occurring, which is $1-e^{-\lambda(1-\theta)}\text{.}$

Again there is no simple closed formula for the overall probability of
management withholding privately observed signals. By way of illustration of
how that probability depends on the intensity $\lambda$ for fixed $\sigma$
and $\kappa$ we show in Fig. 3a a graph, again obtained by using a Bezier spline
through a list of calculated points.

Thus Fig 3a corresponds to a sequence of values of $\lambda $ in the range $%
[1,11]\text{.}$ Here we have computed with 10 points each at distance 1 apart.
Note that again the probability of non-disclosure of a signal observed
below $\gamma _{1}$ tends to $0$ as $\lambda \rightarrow \infty $, since $%
\gamma _{1}$ tends to $0$ as $\lambda \rightarrow \infty .$

\begin{figure}[tbp]
\begin{center}
\includegraphics[height=3.5cm]{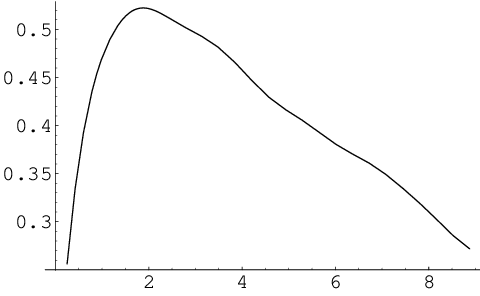}
\par
{Figure 3a. Scienter probability (withholding earlier signals below $%
\gamma_1 $) in Poisson sampled candid-first case with: $\sigma=3,$ $%
\kappa=0.7.$}
\end{center}
\end{figure}

Again for comparison purposes consider the probability that%
\begin{equation*}
\min_{s\in\lbrack0,1-\theta]}\left( -\frac{1}{2}\sigma s+W_{s}\right)
<-\lambda\lbrack g(1)-g(\theta)]/\sigma\text{,}
\end{equation*}
where now the times $s$ are not restricted to be Poisson arrival times. In
this case a closed formula is again available (again, see Appendix A). In
the formula we take $\mu=-\sigma/2$ and $a=-\lambda\lbrack g(1)-g(\theta
)]/\sigma<0\text{,}$ since $g(1)>g(\theta)\text{.}$ The required probability
is:
\begin{equation*}
\Phi\left( \frac{-\lambda\lbrack g(1)-g(\theta)]+\frac{1}{2}\sigma
^{2}(1-\theta)}{\sigma\sqrt{(1-\theta)}}\right)
\end{equation*}
\begin{equation*}
+e^{\lambda\lbrack
g(1)-g(\theta)]}\Phi\left( \frac{-\lambda\lbrack g(1)-g(\theta)]-\frac{1}{2}%
\sigma^{2}(1-\theta)}{\sigma\sqrt{(1-\theta)}}\right) \text{.}
\end{equation*}

Multiplying this by $1-e^{-\lambda (1-\theta )}$ yields the probability of
an observation below $\gamma _{1}\text{.}$ This is graphed in Fig. 3b for a fixed
$\kappa $ against $\lambda \geq \lambda _{\min }^{\mathrm{cand}}\text{.}$

\begin{figure}[tbp]
\begin{center}
\includegraphics[height=3.5cm]{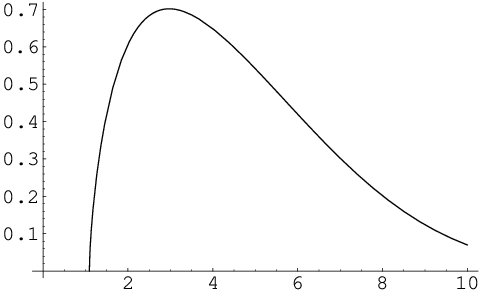}
\par
{Figure 3b. Same withholding probability for continuously monitored
candid-first case with $\sigma=3,\kappa=0.7.$}
\end{center}
\end{figure}

Note that increasing $\kappa$ increases the maximal probability of withheld
observation; increasing $\sigma$ likewise increases this maximum. Here large
$\lambda$ yields a smaller terminal threshold $\gamma_{1}$, whereas small $%
\lambda$ corresponds to lengthier candour.

Notice that this analysis demonstrates that management$\text{ can }$control
cutoffs and corresponding damages-risk by setting $\lambda$ either low
enough (initially offering more candour) or high enough (to lower the
damages cutoff $\gamma_{1}$ from the upper cutoff curve).
\subsection{Legal system policy instruments: interaction}
In this subsection we consider the comparative effect of different legal
systems on equilibrium disclosure behaviour, as modelled by the choices of $%
h(t)$ and $\rho _{t}$ in the framework of (cont-eq) in subsection 3.2. In
particular, we compare and contrast the European and the US systems. The two
systems are equivalent in the special cases of IPOs and other materially
significant contracts. The European system is captured by the choice derived
in subsection 3.2, namely the instantaneous BS-vanilla put:%
\begin{equation*}
h_{\text{vanilla}}(t)=2\Phi (\sigma /2\sqrt{1-t})-1 \qquad (0\leq t \leq 1).
\end{equation*}

To model a significant contrast arising in the US case we introduce a
materiality threshold as does occur in the USA, where only a sufficiently
large valuation fall is open to litigation.
\footnote{%
On April 12, 2024, the US Supreme Court reversed the US Court of Appeals for
the Second Circuit's decision in Macquarie v. Moab Partners and held that a
pure omission cannot form the basis of a securities fraud claim under Rule
10b-5(b). The Supreme Court made clear that an omission is only actionable if it
renders an affirmative statement materially misleading --
US Supreme Court: Pure omissions not actionable under Rule 10b-5(b), Cooley April 22, 2024.}

Specifically, to model this materiality barrier, consider, for simplicity, a fixed threshold
 in which the first $(1-v)\%$ of relative valuation falls do
\textit{not} attract damages. We refer to $1-v$ as the \textit{ materiality standard}, so that the lowest standard is exemplified by $v=1$, for which case $100\%$ of a damage claim is actionable, and a high standard is exemplified by $v=1/10$, say, for which case only the highest decile of claims for valuation falls are actionable. Thus high materiality standards introduce a barrier to smaller damage claims. It emerges that this assumption implies an
amendment to the form of the function $h$ occurring in equation (cont-eq).

We adapt the earlier calculation of $h$ by decomposing the first term in
(cont-eq) with the $v$-threshold%
\begin{align*}
\int_{0}^{\gamma _{t}}(\gamma _{t}-y)d\mathbb{Q}_{t}\mathbb{(}%
y)=\int_{0}^{v\gamma _{t}}(\gamma _{t}-y)d\mathbb{Q}_{t}\mathbb{(}%
y)+\int_{v\gamma _{t}}^{\gamma _{t}}(\gamma _{t}-y)d\mathbb{Q}_{t}\mathbb{(}%
y).
\end{align*}%
The second term is retained whereas the first term combines with the
restorative damages term to yield the following protective put, computed
between time $t$ and time $s$:%
\begin{equation*}
(1-\rho _{t})\int_{0}^{v\gamma _{t}}(\gamma _{t}-y)d\mathbb{Q}_{t}\mathbb{(}%
y)+\int_{v\gamma _{t}}^{\gamma _{t}}(\gamma _{t}-y)d\mathbb{Q}_{t}\mathbb{()}
\end{equation*}%
In order to use standard B-S puts and put-call parity with strike reduced to
$v\gamma _{t}$, a further decomposition is helpful:%
\begin{align*}
&(1-\rho _{t})\int_{0}^{v\gamma _{t}}(v\gamma _{t}-y)d\mathbb{Q}_{t}\mathbb{%
(}y)+(\gamma _{t}-\gamma _{t}v)(1-\rho _{t})\int_{0}^{v\gamma _{t}}d\mathbb{Q%
}_{t}\mathbb{(}y) \\
&+\int_{0}^{\gamma _{t}}(\gamma _{t}-y)d\mathbb{Q}_{t}\mathbb{(}%
y)-\int_{0}^{v\gamma _{t}}(\gamma _{t}v-y)d\mathbb{Q}_{t}\mathbb{(}y)
-\gamma
_{t}(1-v)\int_{0}^{v\gamma _{t}}d\mathbb{Q}_{t}\mathbb{(}y).
\end{align*}%
After some cancellations we have%
\begin{align*}
\int_{0}^{\gamma _{t}}(\gamma _{t}-y)d\mathbb{Q}_{t}\mathbb{(}y)-\rho
_{t}\int_{0}^{v\gamma _{t}}(v\gamma _{t}-y)d\mathbb{Q}_{t}\mathbb{(}y)
-\rho
_{t}\gamma _{t}(1-v)\int_{0}^{v\gamma _{t}}d\mathbb{Q}_{t}\mathbb{(}y).
\end{align*}%
Writing $1_{K}$ for the all-or-nothing binary option with unit payout below $%
K,$ we appeal to put-call parity (in the form $P_{v\gamma }=C_{v\gamma
}+v\gamma _{t}-\gamma _{t})$ to price:%
\begin{align*}
P_{\gamma _{t}}-\rho _{t}P_{v\gamma _{t}}-\rho _{t}\gamma
_{t}(1-v)1_{v\gamma _{t}}.
\end{align*}%
The Black-Scholes valuation is:%
\begin{align*}
\gamma _{t}(\Phi (d_{+}^{1})-\Phi (d_{-}^{1}))
-\rho _{t}[\gamma _{t}\Phi
(d_{+}^{v})-v\gamma _{t}\Phi (d_{-}^{v})+v\gamma _{t}-\gamma _{t}]
-\rho
_{t}\gamma _{t}(1-v)\Phi (d_{-}^{v})
\end{align*}%
with $d_{\pm }^{v}=(-\log v\pm \frac{1}{2}\sigma ^{2}(1-t))/(\sigma \sqrt{1-t%
}).$ This give the protective put as $\gamma _{t}h_{\rho }^{v}(t),$ where%
\begin{align*}
h_{\rho }^{v}(t):=[\Phi (d_{+}^{1})-\Phi (d_{-}^{1})]
+\rho _{t}(1-v)[1-\Phi
(d_{-}^{v})]
-\rho _{t}[\Phi (d_{+}^{v})-v\Phi (d_{-}^{v})].
\end{align*}%
When $v=1,$ this recovers $(1-\rho _{t})\gamma _{t}h_{\rho }^{1}(t)=(1-\rho
_{t})\gamma _{t}h(t).$ Note that for $t=1$ we have $h_{\rho }^{v}(1)=-\rho
_{1}(1-v).$

Thus the different forms of $h(t)$ as we will show generate different $%
\gamma _{t}$ functions and thereby give rise to different equilibrium
disclosure.

The general appearance of the $\gamma _{t}$-dynamic remains in the form%
\begin{equation*}
\gamma _{t}^{\prime }=-\lambda \gamma _{t}h_{\rho }^{v}(t),
\end{equation*}%
and so has solution%
\begin{equation*}
\gamma _{t}=\exp (-\lambda \int_{0}^{t}h_{\rho }^{v}(s)ds).
\end{equation*}%
This reduces to the no-litigation gamma curve under sparing has $\rho =0$ and
$v=1$ and so is%
\begin{equation*}
\gamma _{t}=\exp (-\lambda \int_{0}^{t}h_{\text{vanilla}}(s)ds).
\end{equation*}

The significance of materiality is that it acts in the opposite direction to
the Dye raised damages.
\bigskip

\textbf{Proposition 4(a) (Increase in disclosure with increased materiality standards)} \textit{
In the general case that%
\[
\gamma _{t}^{\prime }=-\lambda \gamma _{t}h_{\rho }^{v}(t),
\]%
as the materiality standard rises (i.e. $1-v$ rises), so the $\gamma _{t}^{v}
$ curve falls, but for $v>0$ the reversal is incomplete:}%
\[
\gamma _{t}^{v}>\gamma _{t}^{0}\equiv\gamma _{t}^{\text{no-lit}}.
\]%
This is self-evident: the larger is $v$, the higher the damages and the lower is materiality  $1-v$. Formally, differentiation of the relevant damages term in the proof of Proposition 1, namely $\displaystyle{\lambda\int_0^{v\gamma_{t}}d\mathbb{Q}_t(y)}$ with respect to $v$ amounts to
$\lambda\gamma_{t}\mathbb{Q}_t(v\gamma_{t})>0$. Thus  increase in materiality  implies decrease in $v$ and hence a fall of the $\gamma_{t}$ curve. As for incompleteness see Prop. 4(b).

We expect $v$ to be a cost to the firm in that small v reduce liability to
only large shocks. So we expect $\theta (v)$ i.e. the timing of switching
out of sparing eventually to recede -- fall back towards $t=0$ as $v$
increases, i.e. switching into candid behaviour comes earlier. However, $%
\theta (v)$ might not be defined for low values of $v$ (i.e. no switching
occurs) and thereafter where it is defined, initially it increases
(extending the sparing interval), but then eventually falls towards $t=0$, as
expected, as illustrated in Figure 4.
\begin{figure}[tbp]
\begin{center}
\includegraphics[height=3.5cm]{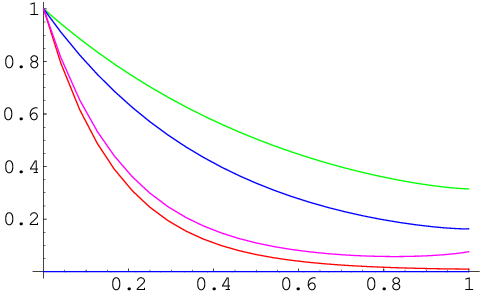}
\par
{Figure 4. At the bottom the no-litigation curve (red).  From the top down curves where litigable claims allow all 100\%, only 80\% and just 5\% of the highest claims. }
\end{center}
\end{figure}

Thus the surprising result obtained with choices of a constant $\rho _{t},$
which we characterize below, is that concentrating on the insurance effect
induced by increased liability omits the important countervailing force of
raising materiality. Only when the joint effect of both policy instruments
interacting is considered can one conclude the aggregate effect on voluntary
disclosure.

After an increase in damages which raises the disclosure threshold
(Proposition 4(b)), the disclosure threshold is lowered if the materiality is
increased, and that induces earlier switching to candid, as illustrated in
Figure 5. This means that if a regulator plans to increase damages $\rho
_{t}$, they should simultaneously take account of the offsetting effects of
varying materiality levels $v$. That is they should co-ordinate the use of
instruments, rather than consider them only separately.

\begin{figure}[tbp]
\begin{center}
\includegraphics[height=3.5cm]{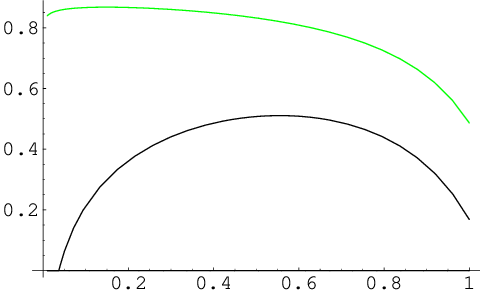}
\par
{Figure 5. Switching-time, where defined, against $v$ for $\sigma=2$ (black) and $\sigma=4$ (green).}
\end{center}
\end{figure}

Whilst our main normative focus in this subsection is upon interacting
policy instruments at the legal framework level, there are also significant
individual effects of the model parameters, namely of firm variability $%
\sigma $, frequency of news arrival $\lambda ,$ and the two
pay-for-performance parameters $\kappa $ and $\beta .$ Thus our model of $%
\theta ,$ the switching time (to candid), provides the following empirical
predictions as shown in the table below (over generic ranges).%
\renewcommand{\arraystretch}{1.25}\newline
\[
\begin{tabular}{|l|l|l|l|}
\hline
$\sigma $ & $\lambda $ & $\kappa $ & $\beta $ \\ \hline
$\theta \uparrow $ & $\theta \uparrow $ & $\theta \uparrow $ & $\theta
\downarrow $ \\ \hline
\end{tabular}%
\]%
\renewcommand{\arraystretch}{1}\newline

In addition to these basic model parameters of the firm, we will draw
attention in the next subsection to legal system effects (which go beyond
the effects of constant $\rho _{t}$ considered in Proposition 2).

\subsection{The litigation charge: incentivizing disclosure}
There are two equivalent approaches a court might follow in adjudicating
damages, one based on liability the other on charges. As these are dual
neither provides a primitive notion.

Depending on the legal system in force, court may assess restitutionary
damages at time $t$ of the form%
\[
\rho _{t}(\gamma _{t}-Y_{t}),
\]%
where the factor $\rho _{t}$ recognizes (partial) scienter liability based
on probable knowledge of bad information (below a given valuation curve $%
\gamma _{s}$ at or before time $t$) which increases with time. Indeed, the
scienter probability (undisclosed knowledge), say $s_{t},$ increases with
time since%
\[
ds_{t}/dt=\lambda \mathbb{Q}_{t}\mathbb{(\gamma }_{t}).
\]

Note that the damages may be applied to a firm on the basis of probable
scienter liability, despite the firm having had no such scienter knowledge.
This may be described as a deadweight that firms may have to pay in the
European legal system. That deadweight may well be much smaller in the
US-system given the imposition of a materiality threshold for claims. This
observation seems new to us. The \textit{deadweight} is given by the formula:

\[
\rho _{t}(\gamma _{t}-Y_{t})\cdot \mathbb{Q}_{t}(Y_{s}<\gamma _{s}\forall
s\leq t\text{ and }\Delta N_{s}>0).
\]

This formula includes the probability that at each Poisson
arrival time $s$ the signal $Y_{s}$ is below the disclosure threshold $%
\gamma _{s}$, so has not been disclosed.

\medskip
Alternatively, subject to the legal system, court may choose to apply
damages in the form of a (penalty) charge $c(t),$ in favour of the
stockholders, reduced according to the relative valuation fall%
\[
c(t)\cdot \frac{\gamma _{t}-Y_{t}}{\gamma _{t}}.
\]%
Here the charge $c(t)$ may possibly increase with time and is weighted
according to the expected asset value $\gamma $, so perhaps of the form%
\[
c(t)=\tau (t)w(\gamma ),
\]%
with $\tau (t)$ an increasing function of $t$ and $w(\gamma )$ a weighting
function responding to the asset value $\gamma .$

The two approaches become equivalent if the damages are identically awarded,
so that
\begin{align*}
\frac{c(t)}{\gamma _{t}}(\gamma _{t}-Y_{t}) &=\frac{\tau (t)w(\gamma _{t})}{%
\gamma _{t}}(\gamma _{t}-Y_{t})=\rho _{t}(\gamma _{t}-Y_{t}): \\
\rho _{t} &=\frac{\tau (t)w(\gamma _{t})}{\gamma _{t}}.
\end{align*}%

Three simple cases come to mind.

(i) $c(t)\equiv c,$ with $0<c<1$ a constant, so that $\rho _{t}=c/\gamma
_{t}.$

(ii) $c(t)\equiv c\cdot t,$ with $0<c<1$ a constant, so that $\rho
_{t}=c\cdot t/\gamma _{t}.$

(iii) $w(\gamma )=\gamma $, so that $\rho _{t}=\tau (t),$ including $\tau
(t)\equiv \rho$ with $0\leq \rho \leq 1$, a constant, and $\tau (t)\equiv c\cdot t,$ with $0<c<1.$

The linear approach mollifies (reduces) the charge initially (for small $t$%
). These may be given further mollification by replacing $ct$ with a power
function $ct^{\nu }$ with $\nu >1$ which downplays low relative valuation
falls and intensifies the charge for relative valuation falls closer to $1.$

The product $c_{t}=\rho _{t}\gamma _{t}$ arising in (i) is
significant in showing $\rho _{t}$ as imposing a charge per unit of
currently expected firm value whereby the damages amount to%
\begin{equation*}
c_{t}\left( \frac{\gamma _{t}-Y_{t}}{\gamma _{t}}\right) .
\end{equation*}%
Written this way the damages term comprises a fraction of the charge $c_t$
determined by the {\it{relative valuation fall}}, i.e.
the ratio of the fall in value to the expected
value.\footnote{If the standardisation $\gamma_0=1$ is dropped, the damages would then have ${\gamma_0}c$ in place of $c$.}
Thus the damages awarded are small when the relative fall is small.
Given a fixed relative fall, a charge $c_t$ increasing with $t$ would reward its earlier disclosure. Henceforth we adopt a constant charge\footnote{%
A fixed charge is a parsimonious choice; the alternative of an increasing $%
c_{t}$ would also force $\rho _{t}$ to increase (as $\gamma _{t}$
decreases), which is key here.} $c$ in order to treat all firms equally (in the same
way), and likewise to treat all disclosure timings equally; we refer to the parameter
$c$ as the \textit{time-invariant} \textit{litigation charge} (constant
relative to expectations), or just \textit{charge} for short. This proposes
a damages mechanism that is normative, independent of time relative to market value, and
furthermore the damages factor $\rho _{t}=c/\gamma _{t}$ is increasing in $%
t. $

The presence of the increasing liability-share factor $\rho _{t}$ above causes the
corresponding equilibrium trend curve $\gamma _{t}$ to rise, just as in the (Dye,
2017) framework. This is not special to this choice, although the situation is particularly transparent: indeed the following result is established in Appendix A.
(The notation $h_{\rho }$ is as in Section 3.4.) The charge $c$ is necessarily below the initial valuation $\gamma_0=1$.
\bigskip

\noindent\textbf{Proposition 4(b) (Damages under sparing raises the dynamic equilibrium
trend value).} \textit{For }$\gamma _{t}$\textit{ the solution to the general equation }$\gamma _{t}^{\prime
}=-\lambda \gamma _{t}h_{\rho }(t),$\textit{ the resulting curve as compared to the
no-litigation curve arising when }$\rho \equiv 0$\textit{ is shifted
upwards by a factor above unity:}

\[
\gamma _{t}=\gamma _{t}^{\text{no-lit}}e^{\lambda j_{\rho }(t)}\text{ where }%
j_{\rho }(t)=\int_{0}^{t}\rho _{u}h(u)du.
\]
\[
\text{So }\rho _{L}<\rho _{H}\quad (\forall t)\Longrightarrow j_{L}<j_{H}\quad (\forall t).
\]
\textit{For the case }$\rho_{t}=\rho_{t}^{c}\equiv c/\gamma_{t}$%
\textit{\ with }$0<c<1,$ \textit{if}%
\begin{equation*}
\gamma_{t}^{\prime}=-\lambda\gamma_{t}h_{\rho}(t)\text{ with }\gamma _{0}=1%
\text{,}
\end{equation*}
\textit{then} $\gamma_{t}$ \textit{is a convex combination of candid and
sparing valuations}:\footnote{%
Note that $\gamma_{t}=c(1-\gamma_{t}^{\text{no lit}})+\gamma_{t}^{\text{no
lit}}$ where $\gamma_{t}^{\text{no lit}}$ represents valuation when
litigation is absent. Thus the option to litigate in the charge framework
amounts to a multiple of the value lost to silence: $1-\gamma_{t}^{\text{no
lit}}.$}%
\begin{equation*}
\gamma^{c}_{t}=c.1+(1-c)e^{-\lambda g(t)}=c(1-e^{-\lambda g(t)})+\gamma_{t}^{\text{no lit}}
\end{equation*}
\begin{equation*}
>\gamma_{t}^{\text{no lit}}=e^{-\lambda g(t)}\text{ (for }t>0\text{).}
\end{equation*}
\textit{Hence }$\partial\gamma_{t}/\partial c>0,$\textit{\ }$\partial\gamma
_{t}/\partial\lambda<0,$ $\partial\gamma_{t}/\partial\sigma<0.$ \textit{%
Furthermore,}%
\begin{equation*}
\rho_{t}^{c}=\frac{c}{c+(1-c)e^{-\lambda g(t)}}=\left[ 1+(c^{-1}-1)e^{-%
\lambda g(t)}\right] ^{-1},
\end{equation*}
\begin{equation*}
\frac{\partial\rho_{t}}{\partial c} =\rho_{t}^{c}c^{-2}e^{-\lambda g(t)}>0,
\end{equation*}
\textit{so that }$\rho_{t}^{c}$\textit{\ increases with }$c.$ \textit{Here
the equilibrium curve }$\gamma_{t}$ \textit{with litigation permitted lies
above the corresponding curve arising when litigation is not permitted and
rises with }$c$\textit{. }\textit{In particular, }$\gamma_{1}^{c}>\gamma_{1}^{\text{no lit}}$.

\section{General litigation effects of switching strategies}

In this section we study the effect that the threat of litigation has on
disclosure strategies involving a switch between candid and sparing
behaviour. Recall that switching is observed emprircally -- see (Grubb, 2011).
We use the setting developed recently in Gietzmann and
Ostaszewski (2023), which we briefly recall. The modelling framework of
activity and observation, determined by a Poisson process of intensity $%
\lambda \text{,}$ is as in the last section. At any time $t\text{,}$
if management have observed a signal $Y_{t}\text{,}$ they may act \textit{%
sparingly} with probability $\pi _{t}$ and disclose $Y_{t}$ on condition
that $Y_{t}$ is above a dynamic threshold $\gamma _{t}$; alternatively, they
may act \textit{candidly} with probability $1-\pi _{t}$ and disclose $Y_{t}$
(without precondition). In the presence of potential litigation with
liability-share\ $\rho _{t}$ the equilibrium valuation satisfies an ODE
similar to (cont-eq) of the preceding section, namely
\begin{align*}
\gamma _{t}^{\prime }=-\lambda \pi _{t}\gamma _{t}h_{\rho }(t)\text{ s.t. }%
\gamma _{0}=1 \tag{$\pi$-Dye}
\end{align*}%
with $h_{\rho }(t)=(1-\rho _{t})h(t)$, where $h(t)$ is as in preceding sections. On intervals where $\pi _{t}=0%
\text{,}$ the valuation therefore does not fall: it remains constant,
providing an improvement relative to the falling valuation of the sparing
mode. Management thus exert some control via $\pi _{t}$ over the valuation
trajectory $\gamma _{t}$ and are assumed to receive a pay-for-performance
reward (which they maximize) in the form%
\begin{align*}
\max_{\pi }\mathbb{E}^{\mathbb{Q}}\int_{0}^{1}(1-\pi _{t})(\alpha \gamma
_{t}-\beta (\gamma _{t}-\gamma _{t}^{1}))\text{ }\mathrm{d}t,  \tag{obj-1}
\end{align*}%
with $\alpha ,\beta $ positive constants related as below. In referring here
to the market probability, this too follows the (Dye, 1985) paradigm which
assumes that management's interests are fully aligned with the
shareholders'. Here, $t=0$ is the most recent time of disclosure and
unit time is left to the next mandatory disclosure (\textit{time to expiry)}%
; $\gamma _{t}^{1}$ denotes the valuation trajectory arising from a
permanently sparing strategy (sparing throughout the interval), i.e the
solution of the ODE\ corresponding to $\pi _{t}\equiv 1.$

This objective includes an instrumental utility\footnote{%
cf. the (Yariv, 2002) analysis of cognitive consistency, separably valuing
actions and beliefs}, a reward proportional to $\gamma _{t}$, and a behavioural
(dis-)utility, a \textit{penalty} proportional to $(\gamma _{t}-\gamma
_{t}^{1}).$ The \textit{amended unravelling principle} (cf. the start of
Section 3.3) implies that
introduction in a market equilibrium of candour (candid reporting) will
cause the valuation $\gamma _{t}$ to exceed $\gamma _{t}^{1}$ and the aim of
the penalty term is to motivate management into protecting the value of the
firm from potential\ falls were a candid strategy followed for too long
(i.e. from excessive use of a candid position under which there is a
likelihood of more frequent downward shocks and more likely reduction of
firm value).

In equivalent form, the objective may be rewritten as
\begin{align*}
\beta \max_{\pi }\mathbb{E}^{\mathbb{Q}}\int\nolimits_{0}^{1}(1-\pi
_{t})[\gamma _{t}^{1}-\kappa \gamma _{t}]\text{{}}\mathrm{d}t  \tag{obj-2}
\end{align*}%
for $\kappa =1-\alpha /\beta$. It is thus natural to demand that for some proper interval of time $\gamma
_{t}^{1}>\kappa \gamma _{t}$ holds, so we make the \textit{blanket assumption%
}
\begin{equation*}
0<\alpha /\beta <1,\text{ i.e. }0<\kappa <1,
\end{equation*}%
which enables discounting of $\gamma _{t}$ by $\kappa $ to a level below $%
\gamma _{t}^{1}.$ For appropriate $\beta ,$ this reward may be traded off
against the expected loss from a settlement arising when $Y_{1}<\gamma _{1},$
namely $\rho _{1}\mathbb{E}^{\mathbb{Q}}[\gamma _{1}-Y_{1}|Y_{1}<\gamma
_{1}],$ yielding an objective with a bequest term:%
\begin{align*}
\max_{\pi }\mathbb{E}^{\mathbb{Q}}\int\nolimits_{0}^{1}(1-\pi _{t})[\gamma
_{t}^{1}-\kappa \gamma _{t}]\text{{}}\mathrm{d}t
\quad -\beta ^{-1}\rho _{1}\mathbb{%
E}^{\mathbb{Q}}[\gamma _{1}-Y_{1}|Y_{1}<\gamma _{1}].  \tag{obj-3}
\end{align*}

We view $\alpha ,\beta $ in addition to $\sigma $ and $\lambda $ as the
firm's parameters. To determine the optimization of $\pi _{t}$, one may
apply the same Hamiltonian analysis as in Gietzmann and Ostaszewski (2023)
to this current objective; it emerges, that almost the same argument as
there leads here to
\bigskip

\noindent\textbf{Theorem (Non-mixing).}\textit{\ A mixing control }$\pi_{t}$%
\textit{\ with }$0<\pi_{t}<1$\textit{\ over any interval of time is
non-optimal.}
\bigskip

That is, singular control is not involved. For a sketch proof see Appendix A.
Thus for a fixed $\lambda $ the optimizing choice of $\pi _{t}$ is piecewise
constant with local value either $0$ or $1$. Assume now at most one
switching time. Its optimal location is identified in Proposition 5 below.
The proof is in Appendix A and generalizes the more special case in Gietzmann
and Ostaszewski (2023). Below, as in Section 3.4,
$g_{\rho }(t)=\int_{0}^{t}h_{\rho }(u)du=\int_{0}^{t}(1-\rho _{u})h(u)du.$
Below the expected loss from silence is denoted by
$$S_1:=\gamma_1-\mathbb{E}^{\mathbb{Q}}[Y_1|Y_1<\gamma_1]. $$
An explicit formula is computed in the proof of the following result.
\bigskip

\noindent \textbf{Proposition 5 (Comparative switching statics). }\textit{%
Assuming a constant liability-share factor $\rho$\ of Section 3.7 (example (iii)) and the objective
function (obj-3), the optimal switching time }$\theta =\theta _{\rho }$%
\textit{\ under a candid-first or sparing-first regime with $h(t)=h_{\theta}(t)$, if such exists, is
given respectively by the first-order conditions:}%
\begin{equation*}
e^{-\lambda g_{\rho }(\theta )}=\kappa,
\end{equation*}%
\textit{and}%
\begin{equation*}
\lambda h_{\rho }(\theta )(1-\theta )=(\kappa ^{-1}-1)+\kappa ^{-1}\gamma
_{\theta }^{-1}\rho _{1}S^{\prime}_{1}(\theta)\beta ^{-1}.
\end{equation*}
\textit{Thus the corresponding switching times satisfy}%
\begin{equation*}
\theta _{\rho }^{\text{candid}}>\theta _{\text{no lit}}^{\text{candid}}\text{
\qquad and \qquad }\theta _{\rho }^{\text{sparing}}<\theta _{\text{no lit}}^{%
\text{sparing}},\text{ }
\end{equation*}%
\textit{i.e. relative to a context which does not permit litigation, under
the threat of litigation candid behaviour lasts for longer. More generally,}%
\begin{align*}
\rho ^{L}_t& <\rho ^{H}_t\text{ }(\forall t)\Longrightarrow \theta _{L}^{%
\text{candid}}<\theta _{H}^{\text{candid}}, \\
\rho ^{L}_t& <\rho ^{H}_t\text{ }(\forall t)\Longrightarrow \theta _{H}^{%
\text{sparing}}<\theta _{L}^{\text{sparing}}.
\end{align*}
\bigskip

\noindent\textbf{Remarks. }1. To obtain a switching equilibrium we need the
two curves involved in the first-order condition above, one rising, one
falling, to intersect. This leads to the following result.
\bigskip

\noindent \textbf{Proposition 6 (A critical charge). }\textit{The sparing-first switching exists for $c<\bar{c}=\bar{c}(\beta),$
where $c=\bar{c}>0$ solves a quadratic equation in $c$. This quadratic has a positive and a negative root if $\lambda h(0)>(\kappa
^{-1}-1),$ a condition which guarantees switching in the no-litigation
model. Under this condition the quadratic is positive at $c=0$ and negative
at $c=1$, so that $0<\bar{c}<1.$}
\bigskip

\noindent2. In the sparing-first circumstance the term $\kappa^{-1}-1$
represents deterrence from over-zealous candid disclosure, the only
surviving term in the absence of litigation (cf. Gietzmann and Ostaszewski
(2023). Here this term is increased by a second term representing litigation
effects so that switching out of sparing behaviour occurs earlier.

\noindent3. In the candid-first circumstance it is the the term $\kappa$
that represents deterrence from over-zealous candid disclosure. Here the
term is decreased by a second term which represents a litigation effect so
that switching out of candid behaviour occurs later.

\noindent 4. In both of the circumstances the deterrence term, be it $\kappa
^{-1}-1$ or $\kappa $, is adjusted. We referred to these adjustments in
converting a constant $\rho $ to the value 0 as in the example of Section
3.5.
\bigskip

\noindent\textbf{Corollary 2.} \textit{For optimal single-switching
strategies applied under sparing behaviour with constant liability-shares }$\rho^{L}<\rho^{H}$ \textit{ :}
\begin{equation*}
\gamma_{1}^{L}<\gamma_{1}^{H}.
\end{equation*}

There is a trade-off to consider between different liability-shares $\rho
_{t}$. Below we introduce the notion of (\textit{incremental})\textit{\
aggregate-disclosure}. For this we need to introduce a monetary measure. We
begin by interpreting the (declining) value of the firm $\gamma
_{t}=e^{-\lambda g_{\rho}(t)}$ (defined as in Section 3.4) as
equivalent to a discount-bond contract offering a continuous coupon of $1$
per unit time, with discount coupon rate $\lambda g_{\rho}(t)$, face
value $\gamma _{0}=1$ (and maturity payment determined at time $t=1$ as $%
Y_{1}$). That is, the continuous analogue of the Dye model generates a
discount coupon rate associated with the valuation of a firm silent until
its terminal mandatory disclosure date.

For $0<\theta <1,$ the coupon rate allows the following integral%
\begin{equation*}
\int_{0}^{\theta }e^{-\lambda g_{\rho}(t)}\text{ }\mathrm{d}t
\end{equation*}%
to be interpreted monetarily as a \textit{bond yield} (or lease valuation),
obtained from holding the bond over the interval $(0,\theta )$. (The
maturity payment is not involved here.) The changing valuation effect of
different liability-share factors which generate different switching times
can then be tracked as a differential bond yield. Thus, for respectively a
High or Low liability-share, the corresponding switching times $\theta _{H}$
and $\theta _{L}$ give rise to different (sparing) `holding times'. This
opens the problem of determining whether investors gain more from a change
in liability-share from Low to High; we achieve this by assessing the
differential voluntary disclosure behaviour as measured by the corresponding
bond yield. There are two bond yields to assess.

One arises as a loss of bond yield (loss of voluntary disclosures) resulting
from the raised liability-share (incremented from Low to High) during the
shortened sparing holding-time under the High charge, namely%
\begin{equation*}
\int_{0}^{\theta _{H}}[e^{-\lambda g_{\rho ^{H}}(t)}-e^{-\lambda g_{\rho
^{L}}(t)}]\text{ }\mathrm{d}t.
\end{equation*}%
This bond yield (area) assesses the lost disclosed values arising in the
interval $[\gamma ^{H}_t,\gamma ^{L}_t]$ for $t<\theta _{H}$ when the
liability-share rises. We call this the \textit{incremental aggregate
voluntary disclosure loss, }or \textit{incremental loss} for short, denoted $%
A_{L}.$

The other assessment accounts for gains in bond yield (voluntary disclosures)
associated with the earlier of the two switching times, namely%
\begin{equation*}
\int_{\theta _{H}}^{\theta _{L}}e^{-\lambda g_{\rho ^{L}}(t)}\text{ }\mathrm{%
d}t.
\end{equation*}%
Here, for each $\theta _{H}<t<\theta _{L},$ disclosed values arise in the
interval $[0,\gamma ^{L}_t]$ in view of the switch to candid behaviour for $%
t>\theta _{H}.$ We call this the \textit{incremental aggregate voluntary
disclosure gain, }or \textit{incremental gain} for short, denoted $A_{G}.$

Unsurprisingly, the bond yield from lowered liability-share (from $c$ to $d$%
) and candid behaviour after the (optimal) switch is generally larger than
the bond yield prior to the switch. An exception to this observation can
occur for small values of the charge $c$ (and small $\beta ,$ so that also $%
\alpha $ is small because $\alpha =(1-\kappa )\beta $). Our Main Theorem
below captures this formally; the notation uses $\theta _{c}$ for the
optimal switching time corresponding to the charge $c$, defined for $c<\bar{c%
}$ as in the remarks following Proposition 5. The parameter $\underline{c}$
is defined at the end of the proof (which is in Appendix A). The two areas $%
A_{L}$ and $A_{G}$ referred to below are illustrated in Figure 6 (for $\beta
=1,\sigma =4)$ and indicate incremental voluntary disclosure when the charge
$c=0.2$ is raised to $d=0.3$, with $\theta _{d}<\theta _{c}.$ Raising the
charge from $c$ to $d$ creates losses $A_{L}$ shown in dark colouring and
gains $A_{G}$ shown in a lighter shade.

\begin{figure}[tbp]
\begin{center}
\includegraphics[height=3.5cm]{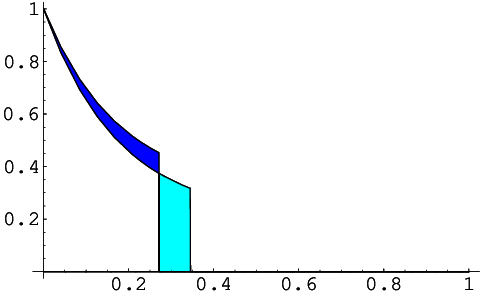}
\par
{Figure 6: Incremental voluntary disclosure against time: areas measure
their aggregate. }
\end{center}
\end{figure}
\bigskip

\noindent \textbf{Main Theorem.} \textit{Assume the liability-share factor
of Section 3.7, the objective function (obj-3) and a standard of materiality $0\leq v \leq 1$. The effect in equilibrium
of incremented (raised) charges on voluntary disclosure behaviour of a large
enough charge }$c>\underline{c}$ \textit{\ in terms of bond yield is that if }
$c<d<\bar{c},$ \textit{ then }%
\begin{align*}
&A_{L}:=\int_{0}^{\theta _{d}}[e^{-\lambda g_{d}(t)}-e^{-\lambda g_{c}(t)}]%
\text{ }\mathrm{d}t \\
< \quad &A_{G}:=\int_{\theta _{d}}^{\theta _{c}}e^{-\lambda
g_{c}(t)}\text{ }\mathrm{d}t.
\end{align*}%
\textit{\ Generally, for small enough }$\beta ,$%
\textit{\ say for }$0<\beta <\underline{\beta },$\textit{\ we have}%
\begin{equation*}
A_{G}>A_{L}\text{ for all }c;
\end{equation*}%
\textit{otherwise, for }$\beta >\underline{\beta }$\textit{\ , there is }$%
\underline{c}=\underline{c}(\beta ),$\textit{\ with }$\underline{c}(\beta )$%
\textit{\ increasing in }$\beta ,$\textit{\ such that}
\begin{align*}
A_{L}& >A_{G}\text{ for all }c<\underline{c}, \\
A_{G}& >A_{L}\text{ for all }c>\underline{c}.
\end{align*}
\textit{In particular, for }$\beta >\underline{\beta }$\textit{, this
implies:}\newline
\noindent \textit{\textbf{Dye-like disclosure reduction: }} \textit{the
incremental (aggregate voluntary disclosure) loss }$A_{L}$ \textit{exceeds
the incremental gain }$A_{G}$\textit{\ in a decreasing manner in the range }$%
0<c<\underline{c}$;\newline
\noindent \textit{\textbf{legal consistency zone: }}(i) \textit{\ the
incremental gain }$A_{G}$ \textit{exceeds }$A_{L}$\textit{\ in an increasing
manner in the range }$\underline{c}<c<\bar{c}$\textit{;}\newline
\noindent and (ii)\textit{\ sparing is non-optimal and candid behaviour is
followed, in the range }$\bar{c}<c\leq 1$.

\textit{Similar results for $1-v$ and $\rho$, with a critical threshold analogous to $\underline{c}$, establish a legal consistency zone for parameters above the critical threshold, namely above the increasing functions $1-\underline{v}(\beta)$, $1-\underline{v}(\lambda)$, and $\underline{\rho}(\lambda)$  and a decreasing  $\underline{\rho}(\beta)$.}
\bigskip

\noindent \textbf{Remark.} Above we refer to the \textit{legal consistency
zone} comprising two circumstances: (i) a switching \textit{%
Goldilocks-interval}, $\underline{c}<c<\bar{c}$, where the litigation charge
that is neither too high nor too low elicits an increase in voluntary
disclosures, together with (ii) a second interval, where candid behaviour is
followed throughout time, both contradicting the Dye disclosure-reduction result.
\bigskip

\noindent \textbf{Sketch proof.} The conclusions of the Main Theorem are not an artifact of the litigation charge and are robust to amending the litigation-sharing function $\rho_{t}$.
Its proof, for which see the Appendix, relies on signing an  expression for the change $\Delta
F=F(c_{L})-F(c_{H})$ in disclosure benefit (transparency) as expressed by an appropriate function $F(c)$ of the charge $c$. That expression depends on the behaviour of just two terms.

This remains the case whatever the
damages parameter $p$, be it $c$ or $\rho$ or $v$, that determines the benefit $F(p),$ and consequently $%
\Delta F,$ as $p$ changes by $\Delta p$. The sign which determines legal
consistency is that of the expression%
\[
\Delta p[\gamma ^{p}(\theta (p))\theta ^{\prime }(p)+\int_{0}^{\theta (p)}%
\frac{\partial }{\partial p}\gamma ^{p}(t)dt].
\]%
We refer to the bracketted two terms as the marginal benefit of voluntary disclosure (determined by a bond yield).

Given a natural order for the cost parameter $p,$ so that larger $p$
increases the damages awarded, the more the switching time $\theta (p)$
recedes towards 0 (as the cost increases) -- candid behaviour comes earlier. Thus the
first term provides a negative effect. The other term is typically positive,
since $\gamma ^{p}(t)$ has exponential (decreasing) format, and increasing
the damages raises the $\gamma ^{p}$-curve (cf. Proposition 4(b)). Hence large enough costs $p$
reduce the positive term and increase the negative term. Thus in variants of
the model one continues to observe that large enough costs are responsible for
improved disclosure behaviour. It is not the product of a particular model
for adjudicating damages, or of a particular model of the damages schedule $R_{t}(.)$.

We note that the disclosure benefit comes at a price: namely the deadweight
arising from a probabilistic settlement of claims suffered by firms that
know that they have not made disclosure because they have not received
any private information (any signals), but are unable to prove this.

\section{Concluding remarks: implications}
           By developing a continuous-time analogue of the traditional two-period voluntary disclosure model with privately observed signals (formally modelled as a martingale under management's filtration), we revisit the well-known Dye paradox that increasing damages awards may reduce aggregate voluntary disclosure. Our analysis identifies two mechanisms that mitigate this result: endogenous switching in disclosure behaviour and the role of materiality standards.

    The continuous-time framework allows management to optimally switch from a sparing disclosure strategy---under which only signals above a dynamic threshold are disclosed---to a candid strategy, in which all privately observed signals are disclosed. This endogenous switching increases aggregate voluntary disclosure. We also show that raising the materiality standard has a qualitatively similar effect by inducing an earlier transition to candid disclosure.

    Our framework models litigation jointly with the choice of materiality standard and, unlike much of the existing literature, does not require the court eventually to observe the withheld private signal (the delayed omniscience assumption). In addition, we allow liability sharing to be either fixed or time-dependent. Under time-dependent liability sharing, earlier disclosure of adverse news attracts proportionally lower damages, creating an additional incentive for timely disclosure.

    Within this setting, our Main Theorem establishes that the Dye paradox need not arise. Instead, we identify a legal consistency zone comprising two distinct regions. In the first, liability parameters are neither too low nor too high, so firms optimally switch from sparing to candid disclosure at an earlier date. In the second, damages are sufficiently high that sparing disclosure is never optimal and firms adopt candid disclosure throughout. In both cases, stronger legal sanctions increase, rather than decrease, voluntary disclosure.

    We further characterize when the legal consistency zone exists by identifying the sign of the marginal benefit of voluntary disclosure, which in our framework is determined by an equilibrium bond-yield condition. From an empirical perspective, our principal implication is that a legal consistency zone is more likely to arise when the model's key regulatory parameters---the damages rate, the liability-sharing factor, and the materiality standard (($c$,$\rho$,$(1-v)$)---are sufficiently high. More broadly, our results suggest that damages and materiality standards should be viewed as complementary policy instruments. Their interaction, rather than either instrument in isolation, determines whether litigation encourages or discourages voluntary disclosure.



\section*{References}
\quad Acharya, V., DeMarzo, P. and Kremer I. 2011. \textquotedblleft
Endogenous information flows and the clustering of
announcements.\textquotedblright\ \textsl{Amer. Econ. Rev.} \textbf{101},
2955-79

Bingham, N. H. and Kiesel, R. 1998. \textsl{Risk-neutral valuation.}
Springer.

Bj\"{o}rk, T. and Murgoci, A. 2014. \textquotedblleft A theory of Markovian
time-inconsistent stochastic control in discrete time.\textquotedblright\
\textsl{Finance Stoch.} \textbf{18}, 545--592.

Choi, Albert H., and Kathryn E. Spier. 2022. \textquotedblleft Liability for
Non-Disclosure in Equity Financing. Available\textquotedblright\ at SSRN
4056233.

Cornell, B., and Morgan, R. G. 1989. \textquotedblleft Using finance theory
to measure damages in fraud on the market cases.\textquotedblright\ \textsl{%
UCLA L. Rev.} \textbf{37}, 883.

Daley, D. J. and Vere-Jones, D. 2003. \textsl{An Introduction to the Theory
of Point Processes}, Vol. I, 2nd edn. Springer, 2003.

Daley, D. J. and Vere-Jones, D. 2008. \textsl{An Introduction to the Theory
of Point Processes}, Vol. II, 2nd edn. Springer.

Dybvig, P., Gong, N., and Schwartz, R. 2000. \textquotedblleft Bias of
Damage Awards and Free Options in Securities Litigation.\textquotedblright\
\textsl{Journal of Financial Intermediation} \textbf{9}, 149--68.

Dye, R. A. 1985. \textquotedblleft Disclosure of Nonproprietary
Information.\textquotedblright\ \textsl{Journal of Accounting Research}
\textbf{23}, 123-145.

Dye, R. A. 2017. \textquotedblleft Optimal disclosure decisions when there
are penalties for nondisclosure.\textquotedblright\ \textsl{RAND Journal of
Economics} \textbf{48} no. 3, 704-732.

Dye, R. A., and Sridhar S. S. 2024. \textquotedblleft Management forecasts:
Biases, incentives, and spillover effects.\textquotedblright\ \textsl{RAND
Journal of Economics} 55 no. 4.

Feldman, D. 1992. \textquotedblleft Logarithmic preferences, myopic
decisions, and incomplete information.\textquotedblright\ \textsl{J.
Financial Quant. Anal.} \textbf{27}, 619--629.

Gietzmann, M. B. and Ostaszewski, A. J. 2016. \textquotedblleft The sound of
silence: equilibrium filtering and optimal censoring in financial
markets.\textquotedblright\ \textsl{Adv. in Appl. Probab.} \textbf{48A},
119--144.

Gietzmann, M. B. and Ostaszewski, A. J. 2023. \textquotedblleft The Kind of
Silence: Managing a Reputation for Voluntary Disclosure in Financial
Markets.\textquotedblright\ \textsl{Annals of Finance}, \textbf{19}, 419-447.

Grubb, M. D. 2011. \textquotedblleft Developing a Reputation for
Reticence.\textquotedblright\ \textsl{J. Econ. Man. Strat.} \textbf{20},
225-268.

Jung, W.-O. and Kwon, Y. K. 1988. \textquotedblleft Disclosure when the
market is unsure of information endowment of managers.\textquotedblright\
\textsl{J. Accounting Res.} \textbf{26}, 146--153.

Karatzas, I. and Shreve, S. 2014. \textsl{Brownian motion and stochastic
calculus}, Springer.

Mnookin, R.H, 2021. Bargaining in the shadow of the law Reassessed, in:
Hinshaw A. et al. (eds). \textsl{Discussions in Dispute Resolution: The
Foundational Articles}, 22-26. Oxford University Press.

Mosca, C. and Picciau, C., How EU and U.S. Disclosure Requirements Differ While Sharing the Same Goals
 May 23, 2023, Columbia Law School Blue Sky Blog.

Ostaszewski, A. J. and Gietzmann, M. B. 2008. \textquotedblleft Value
creation with Dye's disclosure option: optimal risk-shielding with an upper
tailed disclosure strategy.\textquotedblright\ \textsl{Review of
Quantitative Finance and Accounting} \textbf{31}, 1-27.

Schantl, S. F., and Wagenhofer, A. 2024. "Economic effects of litigation
risk on corporate disclosure and innovation." \textsl{Review of Accounting
Studies} \textbf{29}, no. 4, 3328-3368.

Yariv, L. 2002. \textquotedblleft I'll See It When I Believe It --- A Simple
Model of Cognitive Consistency\textquotedblright\ \textsl{Cowles Foundation
Discussion Papers. }\textbf{1616}.

\section*{Appendix A: Proofs}

\noindent \textbf{Proof of Proposition 1. }To pass to our continuous-time
setting, rearrange (cond) as follows.%
\begin{align*}
\tilde{q}\gamma _{s}+q\gamma _{s}\int\nolimits_{0}^{\gamma _{s}}d\mathbb{Q}_{t}(y)
&=\tilde{q}\gamma _{t}+q\int\nolimits_{0}^{\gamma _{s}}yd\mathbb{Q}_{t}(y)
+q\int\nolimits_{0}^{\gamma _{s}}\gamma _{s}R_{s}(1-y/\gamma_{s})d\mathbb{Q}_{t}(y)\\
&=\tilde{q}\gamma _{t}+q\int\nolimits_{0}^{\gamma _{s}}yd\mathbb{Q}_{t}(y)
+q\int\nolimits_{0}^{\gamma _{s}}R^{\prime}_{s}(1-y/\gamma_{s})\mathbb{Q}_{t}(y)dy,
\end{align*}
where the last term has been integrated by parts, so that the prime indicates differentiation w.r.t. $y$. Now we apply the mean-value theorem for integrals to obtain for some $y^{*}_{s} \in(0,\gamma_{s})$
\begin{align*}
&q\int\nolimits_{0}^{\gamma _{s}}R^{\prime}_{s}(1-y/\gamma_{s})\mathbb{Q}_{t}(y)dy\\
&=qR^{\prime}_{s}(1-y^{*}_{s}/\gamma_{s})\int\nolimits_{0}^{\gamma _{s}}\mathbb{Q}_{t}(y)dy\\
&=q\rho^{*}_{s}\int\nolimits_{0}^{\gamma _{s}}\mathbb{Q}_{t}(y)dy,
\end{align*}
where we have written $\rho^{*}_{s}$ for $R^{\prime}_{s}(1-y^{*}/\gamma_{s})$.
With the Poisson process in place, for $t<s$ take $q=q_{ts}=\lambda
(s-t)+o(s-t)\text{,}$ employing the Landau little-oh notation. Reassembling terms, passage to
the limit as $s\searrow t$ yields:%
\begin{align*}
&(1-q_{ts})(\gamma_{t}-\gamma_{s})+o(s-t) \\
&= q_{ts}\int\nolimits_{y\leq\gamma_{s}}(y-\gamma_{s})\text{{}}%
\mathrm{d}\mathbb{Q}_{t}(y)\\
&+q_{ts}\rho^{*}_{s}\int\nolimits_{0}^{\gamma _{s}}\mathbb{Q}_{t}(y)dy\\
&= -\lambda(s-t)(1-\rho^{*}_{s})\int\nolimits_{y\leq\gamma_{s}}\mathbb{Q}%
_{t}(y)\text{{}}\mathrm{d}y.\text{ (By parts.)}
\end{align*}

Above, $\mathbb{Q}_{t}((\gamma _{s}-Y_{s})^{+})$ is a protective put. Proceed hereafter as in Section 3.2. \hfill $\square $

\bigskip

\noindent \textbf{Proof of Proposition 3 and Corollary 1. }Throughout this
proof, conditioning is exceptionally under $\mathcal{Y}_{t},$ the
management's private information, and we use $\mathbb{E}_{t}^{\mathcal{Y}%
}[.] $ to mean $\mathbb{E}^{\mathbb{Q}}[.|\mathcal{Y}_{t}].$ We note first
an inequality involving a `conditional call':
\begin{equation*}
\mathbb{E}_{t}^{\mathcal{Y}}[(Y_{1}-\gamma _{1})^{+}|Y_{t}=y]+\gamma _{1}>y%
\text{.}
\end{equation*}%
For a proof when $y<\gamma _{1}$, see Appendix A.

If $Y_{1}<\gamma _{1}$ (at time $t=1)$ the firm suffers damages of $\mathbb{%
\rho }(\gamma _{1}-Y_{1})^{+}$ with the \textit{liability-share } $0<\mathbb{%
\rho }\leq 1$ independent of $Y_{t},$ but dependent on how long the sparing
behaviour has been followed. Recall that the firm value at time $t=1$
becomes
\begin{align*}
Y_{1}-\mathbb{\rho }(\gamma _{1}-Y_{1})^{+}=\left\{
\begin{array}{cc}
(1+\mathbb{\rho })\left[ Y_{1}-\frac{\mathbb{\rho} \gamma_{1}}{1+\mathbb{\rho }}\right]: & \text{}Y_{1}\leq \gamma _{1} \\
Y_{1}: & \text{}Y_{1}>\gamma _{1}\text{.}%
\end{array}%
\right.  \tag{Net-v-1}
\end{align*}%
Here, at first, we will allow the firm to go into debt at $t=1;$ otherwise,
bankruptcy occurs if $Y_{1}<\bar{y}_{1}:=\gamma _{1}\rho /(1+\rho )<\gamma
_{1}$. So, if at time $t$ the observation $Y_{t}$ is received, consider the
conditional expected valuation at time $t=1$ net of damages which is
\begin{align*}
&\mathbb{E}_{t}^{\mathcal{Y}}[Y_{1}-\rho (\gamma
_{1}-Y_{1})^{+}|Y_{t}]\\
&\quad = Y_{t}-\rho \mathbb{E}_{t}^{\mathcal{Y}}[(\gamma
_{1}-Y_{1})^{+}|Y_{t}]\text{.}
\end{align*}%
Exploiting put-call parity and rearranging,
\begin{equation*}
\mathbb{E}_{t}^{\mathcal{Y}}[(\gamma _{1}-Y_{1})^{+}|Y_{t}]=\gamma
_{1}-Y_{t}+\mathbb{E}_{t}^{\mathcal{Y}}[(Y_{1}-\gamma _{1})^{+}|Y_{t}]\text{.%
}
\end{equation*}%
Thus the net-of-damages expected terminal valuation when $Y_{t}$ is observed
is
\begin{align*}
&Y_{t}-\rho \mathbb{E}_{t}^{\mathcal{Y}}[(\gamma
_{1}-Y_{1})^{+}|Y_{t}]\\
&\quad =(1+\rho )Y_{t}-\rho \gamma _{1}-\rho \mathbb{E}_{t}^{%
\mathcal{Y}}[(Y_{1}-\gamma _{1})^{+}|Y_{t}]\text{.}
\end{align*}%
An indifference between disclosure and non-disclosure given $Y_{t}=y$ would
require%
\begin{align*}
y&=(1+\rho )y-\rho \gamma _{1}-\rho \lbrack \mathbb{E}_{t}^{\mathcal{Y}%
}(Y_{1}-\gamma _{1})^{+}|Y_{t}=y],\\
y&=\gamma _{1}+\mathbb{E}_{t}^{%
\mathcal{Y}}[(Y_{1}-\gamma _{1})^{+}|Y_{t}=y]\text{.}
\end{align*}%
However, this equality never holds, because as noted above
\begin{equation*}
y<\gamma _{1}+\mathbb{E}_{t}^{\mathcal{Y}}[(Y_{1}-\gamma _{1})^{+}|Y_{t}=y]%
\text{.}
\end{equation*}%
This last yields, on multiplying by $\rho $ and adding $y\text{,}$
\begin{equation*}
(1+\rho )y<y+\rho \gamma _{1}+\rho \mathbb{E}_{t}^{\mathcal{Y}%
}[(Y_{1}-\gamma _{1})^{+}|Y_{t}=y]\text{ i.e.}
\end{equation*}
\begin{equation*}
(1+\rho )y-\rho \{\gamma
_{1}+\mathbb{E}_{t}^{\mathcal{Y}}[(Y_{1}-\gamma _{1})^{+}|Y_{t}=y]\}<y\text{.%
}
\end{equation*}%
So the firm must choose between maintaining its valuation $\gamma _{t}$
while risking non-disclosure litigation versus disclosing the signal $%
Y_{t}=y<\gamma _{1}$:
\begin{equation*}
\text{conditional net-of-damages valuation}_{t}
\end{equation*}
\begin{equation*}
=(1+\rho )Y_{t}-\rho
(\gamma _{1}+\mathbb{E}_{t}^{\mathcal{Y}}[(Y_{1}-\gamma
_{1})^{+}|Y_{t}])<Y_{t}\text{.}
\end{equation*}

So at time $t$ management expect the value at time $1$ to fall below the
private current forecast $Y_{t}.$

This implies the lower boundary for non-disclosure remains at or above $%
\gamma _{1}\text{,}$ so that $\delta _{t}\equiv \gamma _{1}\text{.}$

Of course,
\begin{equation*}
Y_{t}<\bar{y}_{1}=\gamma _{1}\rho /(1+\rho )\text{, implies }Y_{t}=\mathbb{E}%
_{t}^{\mathcal{Y}}[Y_{1}]<\bar{y}_{1}\text{,}
\end{equation*}%
so that \textit{post-damages bankruptcy} occurs and the value of the firm
will be zero. In this case management is indifferent to disclosure, except
that personal liabilities might enforce disclosure.\hfill $\square $
\bigskip

\noindent \textbf{Proof of Proposition 4(b). }
Solving for $\gamma $ and splitting the integral, here we have:%
\begin{align*}
\gamma (t)&=\exp \left( -\lambda \int_{0}^{t}(1-\rho _{u})h(u)du\right)\\
&=e^{-\lambda g(t)}e^{\lambda j_{\rho }(t)},
\text{ where }j_{\rho}(t)=\int_{0}^{t}\rho _{u}h(u)du.
\end{align*}%
Hence $\gamma (t) = \gamma _{t}^{\text{no-lit}}e^{\lambda j_{\rho }(t)}$ where
\begin{align*}
j^{\prime }(t) &=\rho _{t}h(t)=\rho _{t}\left( 2\Phi (\sigma
/2\surd (1-t))-1\right) >0.
\end{align*}

Here $\rho _{t}$ is increasing while $h(t)$ decreases down to zero, so that
the loading factor $e^{\lambda j_{\rho }(t)}$ is flat at $t=1.$

Higher liability-share factors raise the $\gamma $-curve more, since%
\[
\rho _{L}<\rho _{H}\Longrightarrow j_{L}<j_{H},
\]%

For the constant charge case, write $\gamma $ for $\gamma _{t}$
so that $\rho =c/\gamma .$ Since $\rho ^{\prime }dt=d\rho ,$
\begin{align*}
\rho ^{\prime } =-c\gamma ^{\prime }/\gamma ^{2}&=\lambda (1-\rho )\gamma
hc/\gamma ^{2}\\
&=\lambda (1-\rho )\rho h \\
\int \frac{\text{ }\mathrm{d}\rho }{\rho (1-\rho )} &=\int \left( \frac{1}{%
\rho }-\frac{1}{\rho -1}\right) \text{ }\mathrm{d}\rho \\
&=\lambda \int h\text{
}\mathrm{d}t=\lambda g(t)+\text{ const.}
\end{align*}
\begin{align*}
&\log \rho -\log (\rho -1) =\lambda g(t)+k\text{ (say)} \\
&\frac{\rho -1}{\rho }=1-\frac{1}{\rho }=ae^{-\lambda g(t)}\text{ with }%
a=e^{-k}.
\end{align*}%
Thus $\displaystyle{\gamma /c=1-ae^{-\lambda g(t)}}.$
Taking $t=0$ gives (since $\gamma _{0}=1$) $a=1-c^{-1}=(c-1)/c.$
Substitution gives
\begin{equation*}
\gamma _{t}=c(1-ae^{-\lambda g(t)})=c+(1-c)e^{-\lambda g(t)}>e^{-\lambda
g(t)},
\end{equation*}%
being a convex combination (as $0<c<1$). Note that%
\begin{equation*}
\gamma _{1}=c(1-e^{-\lambda g(1)})+e^{-\lambda g(1)}
\end{equation*}%
so again this being a convex combination of $1$ and $c$ is itself between $c$
and $1.$ Thus%
\begin{equation*}
\rho _{t}=\frac{c}{c+(1-c)e^{-\lambda g(t)}}
\end{equation*}%
and rearrangement yields both the first claim and that $\rho ^{c}$ increases
with $c.$
Since $\gamma _{t}=c+(1-c)e^{-\lambda g(t)},$ this last and the
observation that%
\begin{align*}
&\frac{\partial }{\partial \sigma }g(t;\sigma ) =\int_{0}^{t}\frac{\partial
}{\partial \sigma }h(u;\sigma )\text{ }\mathrm{d}u \\
&=\int_{0}^{t}\frac{%
\partial }{\partial \sigma }[2\Phi (\sigma /2\sqrt{1-u})-1]\text{ }\mathrm{d}%
u \\
& =\int_{0}^{t}[\varphi (\sigma /2\sqrt{1-u})\sqrt{1-u}]\text{ }\mathrm{d}%
u>0,
\end{align*}%
combine to give $\partial \gamma _{t}/\partial \lambda <0$ and $\partial
\gamma _{t}/\partial \sigma <0$ directly.\hfill $\square $

\bigskip

\noindent\textbf{Remark 1. }The litigation-inclusive curve lies above the
litigations-`prohibited' curve. Indeed, the gamma curve in both situations
is monotonically decreasing, so it is enough to compare the lowest points.

\noindent \textbf{Remark 2. }We may view $\rho $ and $\gamma $ as images
under involution with opposite rates of growth so arising from an identical
kernel, modulo sign $
-\rho ^{\prime }/\rho =-\lambda (1-\rho )h=\gamma ^{\prime }/\gamma \text{.}$
\bigskip

\noindent \textbf{Proof of the Non-mixing Theorem.} Assume otherwise. The
same calculations as for Theorem S1 in Gietzmann and Ostaszewski (2023) lead
to the conclusion that $1/h_{\rho }(t)=\lambda t+k.$ Hence, substituting $%
1-\rho _{t}=1-c/\gamma _{t}$ below yields for all, $t\in
\lbrack 0,1],$
\begin{align*}
1/((k+\lambda t)h(t))& =(1-\rho _{t})\\
&=1-\frac{c}{c(1-e^{-\lambda g(t)})+e^{-\lambda g(t)}}, \\
(k+\lambda t)h(t)& =1+e^{\lambda g(t)}(c/(1-c)).
\end{align*}%
But $h(t)=2\Phi (\hat{\sigma}/2)-1$ with $\hat{\sigma}=\sigma (1-t),$ which
gives a contradiction.\hfill $\square $

\bigskip

\noindent \textbf{Proof of Proposition 5. }We begin with computation of
the \textit{expected value below}\textbf{\ }$\gamma _{1}.$ Working
in the standard setting of sparing behaviour over $[0,1],$ the event $%
(Y_{1}<\gamma _{1})$ corresponds to%
\begin{align*}
\log Y_{1}&=-\frac{1}{2}\sigma ^{2}+\sigma W_{1}<\log \gamma _{1},\text{ so
that }\\
W_{1}&<\eta :=\frac{1}{2}\sigma +(\log \gamma _{1})/\sigma ,
\end{align*}%
where $\gamma _{1}$ is computed under the litigation mechanism running over $%
[0,1]$. The required expectation $\mathbb{E}_{0}^{\mathbb{Q}%
}[Y_{1}|Y_{1}<\gamma _{1}]$ is as below, with $\varphi (w):=e^{-\frac{1}{2}%
w^{2}}/\sqrt{2\pi }:$
\begin{align*}
&\int_{-\infty }^{\eta }\exp (-\frac{1}{2}\sigma ^{2}+\sigma w)\varphi
(w)dw\\
 &=\int_{-\infty }^{\eta }\varphi (w-\sigma )dw =\Phi (\eta -\sigma ).
\end{align*}

With a switch at $\theta $ from candid to sparing behaviour,
there is no litigation possible in $(0,\theta ).$ So, at $\theta $ when
sparing behaviour begins, we have effectively a reset of the origin for the
dynamics. The expected loss from silence over $(\theta ,1)$ may be computed
\textit{after a change of origin and scale} to the standard set-up over $%
(0,1).$

We now consider the optimal choice of switching time $\theta.$

Consider the candid-first case. Since $\gamma _{t}\equiv 1$ on $(0,\theta )$
the objective reduces to:%
\begin{equation*}
\int_{0}^{\theta }e^{-\lambda g_{\rho }(t)}dt-\kappa \theta -\beta ^{-1}\rho
_{1}^{\theta }\mathbb{E}_{0}^{\mathbb{Q}}[\gamma _{1}^{\theta }-Y_{1}|Y_{1}<\gamma
_{1}^{\theta }],
\end{equation*}%
where $\gamma _{t}^{\theta },\rho _{1}^{\theta }$ etc. denotes the
trajectory for $t\in (\theta ,1)$ with $\gamma _{\theta }^{\theta }=1$ and $%
\rho _{1}^{\theta }=c(\gamma _{t}^{\theta })^{-1}.$
We compute the expectation after making a change of origin and scale taking $%
(\theta ,1)$ to $(0,1),$ so that $\rho _{1}$ replaces $\rho _{1}^{\theta }.$
Thus
for $t\in \lbrack \theta ,1]$ we have%
\[
h_{\theta }(t):=h((t-\theta )/(1-\theta )),
\]%
so that $\displaystyle{\gamma ^{\theta }(t)=\exp [-\lambda (1-\rho )g_{\theta }(t)]}$,\text{ where }%
$\displaystyle{g_{\theta }(t)=g((t-\theta )/(1-\theta ))}$
and in particular $\log \gamma _{1}=-\lambda (1-\rho )g(1),$
independently of $\theta .$  Optimizing $\theta $ from%
\begin{align*}
&\int_{0}^{\theta }\gamma ^{1}dt-\kappa \theta -\rho S_{1}\beta
^{-1}\\
&=\int_{0}^{\theta }e^{-\lambda (1-\rho )g(t)}dt-\kappa \theta -\rho
S_{1}\beta ^{-1}
\end{align*}%
yields the F.O.C.:%
\[
e^{-\lambda (1-\rho )g(\theta )}=\kappa :\qquad g(\theta ):=(-\log \kappa
)/(1-\rho ).
\]%
As $h>0$ and $1-\rho_t <1,$ here $h_{\rho }(t)=(1-\rho _{t})h(t)<h(t)$ and so $%
g_{\rho }(t)<g(t)$ and so $e^{-\lambda g_{\rho }(t)}>e^{-\lambda g(t)}$. So $%
\theta _{\rho }>\theta ^{\text{no lit}}$.

Similarly: if $0<\rho ^{L}_t<\rho ^{H}_t<1$ on $(0,1),$ then $1-\rho
^{H}_t<1-\rho ^{L}_t,$ so%
\begin{align*}
h_{\rho }^{H}(t)&=(1-\rho _{t}^{H})h(t)<h_{\rho }^{L}(t)=(1-\rho
_{t}^{L})h(t):\\
 & g_{\rho }^{H}<g_{\rho }^{L}:\qquad e^{-\lambda g_{\rho
}^{L}(\theta )}<e^{-\lambda g_{\rho }^{H}(\theta )}:\\
&\rho ^{L}_t<\rho ^{H}_t\Longrightarrow \theta _{\text{candid}}^{L}<\theta
_{\text{candid}}^{H},
\end{align*}%
in particular, taking $\rho ^{L}=0$ we obtain $\theta _{\text{candid}}^{\rho
}>\theta _{\text{candid}}^{\text{no lit}}.$

Consider the contrasting sparing-first case where $\gamma _{t}\equiv \gamma _{\theta }$
for $t>\theta .$ The computation of the expected loss from silence over $%
(0,\theta )$ may again be standardized by scaling this $\theta $-lengthened
interval to $(0,1)$ and so the objective reduces to
\begin{equation*}
\int_{\theta }^{1}\gamma _{t}^{1}\text{ }\mathrm{d}t-\kappa \gamma _{\theta
}(1-\theta )-\theta \rho _{1}S_{1}(\theta)\beta ^{-1}.
\end{equation*}%
Recall that $\gamma _{t}^{1}=e^{-\lambda g_{\rho }(t)}=\gamma _{t}$ for $%
t<\theta $ so that $\gamma _{\theta }^{1}=\gamma _{\theta }$ and
\begin{align*}
S_{1}(\theta )&=\gamma _{1}-\Phi (\eta -\sigma )=\exp (-\lambda (1-\rho
)g(\theta ))\\
&-\Phi (-\sigma /2-\lambda (1-\rho )g(\theta )/\sigma ).
\end{align*}%
The FOC here is%
\begin{align*}
0=-\gamma _{\theta }+\kappa \gamma _{\theta }&+\kappa \lambda \gamma _{\theta
}h_{\rho }(\theta )(1-\theta )-\rho _{1}S^{\prime}_{1}(\theta)\beta ^{-1}:\\
\lambda h_{\rho }(\theta )(1-\theta )&=(\kappa ^{-1}-1)+(\beta \kappa
)^{-1}\cdot \gamma _{\theta }^{-1}\rho _{1}S^{\prime}_{1}(\theta).
\end{align*}
Here again we have $h_{\rho }(t)(1-t)=(1-\rho _{t})h(t)(1-t)<h(t)(1-t)$ and
so $\theta _{\rho }<\theta ^{\text{no lit}}.$

Here $h_{\rho }^{H}(t)(1-t)<h_{\rho }^{L}(t)(1-t)$ so%
\begin{equation*}
\rho _{t}^{L}<\rho _{t}^{H}\Longrightarrow \theta _{\text{sparing}%
}^{H}<\theta _{\text{sparing}}^{L}.
\end{equation*}%
In particular, taking $\rho ^{L}=0$ we obtain $\theta _{\text{sparing}%
}^{\rho }<\theta _{\text{sparing}}^{\text{no lit}}.$\hfill $\square $
\bigskip

\textbf{Remarks }1. In the sparing-first case, after the switching time $%
\theta ,$ we have $\gamma _{t}\equiv \gamma _{\theta }$ for $t\geq \theta .$
Here $\gamma _{1}=\gamma _{\theta }.$ Continued silence followed by a drop $%
Y_{1}$ leads to the settlement $\rho _{\theta }(\gamma _{\theta }-Y_{1}).$
If instead there is a signal received at time $\tau \in (\theta ,1),$ this
is disclosed and the settlement is $\rho _{\tau }(\gamma _{\theta }-Y_{\tau
})$ and so for consistency we require that $\rho _{t}\equiv \rho _{\theta }$
for $t>\theta $ (in the candid interval).
\bigskip

2. Now consider the candid-first scenario with continued silence. In the
candid period, non-disclosure is interpreted by the market as meaning no new
information and so $\gamma _{t}\equiv 1$ for $t$ in $(0,\theta ).$ The
subsequent period is sparing and again the damages settlement is $\rho _{1}(\gamma
_{1}-Y_{1}).$ Here the dynamics for $t>\theta $ are computed by shifting the
origin and rescaling so that $\log \gamma_1=\lambda (1-\rho)g(1)$.

If the firm were to receive a signal $Y_{\tau}$ at some time $\tau>\theta$
and its disclosure yields  damages of%
\begin{equation*}
\bar{\rho}_{\tau}(\gamma_{\tau}-Y_{\tau})\text{ for }Y_{\tau}<\gamma _{1}%
\end{equation*}
where $\bar{\rho}_{t}=\rho_{(t-\theta)/(1-\theta)\text{ }}\text{for }%
t>\theta$. Thus for an equilibrium to occur we must have%
\begin{align*}
&\int_{0}^{\theta}e^{-\lambda(1-\rho_{t})g(t)}dt-\kappa\theta >\\
&\rho_{1}^{%
\theta
}(\gamma_{1}^{\theta}-E[Y_{1}|Y_{1}<\gamma_{1}^{\theta}])/\beta=(1-\theta
)\rho_{1}^{\text{standard}}S_{1}/\beta.
\end{align*}

\noindent \textbf{Proof of Proposition 6 }
At the origin the falling curve must thus be located
above the rising one, i.e.%
\begin{equation*}
\lambda(1-\rho_{0})h(0)>(\kappa^{-1}-1)+\rho_{0}\rho_{1}(%
\kappa^{-1}S_{1}c^{-1}\beta^{-1}),
\end{equation*}
equivalently, since $\rho_{0}=c,$%
\begin{align*}
&\lambda(1-c)h(0)\\
&> (\kappa^{-1}-1)+\frac{c}{c+(1-c)e^{-\lambda g(t)}}%
(\kappa^{-1}S_{1}\beta^{-1}),
\end{align*}
after substituting for $\rho_{1}$ from Proposition 4(b). That is, we require
\begin{align*}
&\lambda h(0)(1-c)[c+(1-c)e^{-\lambda
g(t)}]\\
&>(\kappa^{-1}-1)[c+(1-c)e^{-\lambda
g(1)}]+c(\kappa^{-1}S_{1}\beta^{-1}).
\end{align*}
Thus the sparing-first switching exists for $c<\bar{c}=\bar{c}(\beta),$
where $c=\bar{c}>0$ solves:%
\begin{align*}
0 & =\lambda h(0)[c^{2}(e^{-\lambda g(1)}-1)+c(1-2e^{-\lambda g(1)})] \\
&-c(\kappa^{-1}-1)(1-e^{-\lambda g(1)}) \\
&-c(\kappa^{-1}S_{1}\beta
^{-1})+[\lambda h(0)-(\kappa^{-1}-1)]e^{-\lambda g(1)}
\end{align*}
This quadratic in $c$ has a positive and a negative root if $\lambda h(0)>(\kappa
^{-1}-1),$ a condition which guarantees switching in the no-litigation
model. Under this condition the quadratic is positive at $c=0$ and negative
at $c=1$, so that $0<\bar{c}<1.$ That is, for $1\geq c>\bar{c}$ only candid
behaviour is optimal. It may be checked that $\bar{c}(\beta)$ is increasing
in $\beta$.
\bigskip

\noindent\textbf{Proof of Corollary 2.} Here by Proposition 5, $\theta_{H}<\theta_{L}.$
Integrating as in Section 3.4 from $0$ to $t\leq \theta_{H}$ equation ($\pi$-Dye) with $\pi_t=0$
gives (since $\gamma_0=1$)
\begin{align*}
\log\gamma_{t}^{H}&=\int_0^t(1-\rho_{s}^{H})h(s)ds\\
&\leq \int_0^t(1-\rho_{s}^{L})h(s)ds =\log\gamma_{t}^{L}
\end{align*}
as $\rho_{t}^{H}\geq\rho_{t}^{L}$.%
\text{ }So, as $\gamma_{t}$ is
decreasing, the initial comparison gives%
\begin{equation*}
\log\gamma_{\theta^{L}}^{L}<\log\gamma_{\theta^{H}}^{L}<\log\gamma_{\theta
^{H}}^{H}=\log\gamma_{\theta^{L}}^{H}.
\end{equation*}
Thus we have $\gamma_{1}^{L}<\gamma_{1}^{H}.$\hfill $\square $
\bigskip

\noindent \textbf{Proof of Main Theorem.} Since $g^{\prime }(t)=h(t)$ and
recalling $\gamma^{c}$ from Prop. 4%
\begin{align*}
\gamma^{c}& =c+(1-c)e^{-\lambda g(t)}:\\
\frac{\partial }{\partial c%
}\gamma^{c}(t)&=1-e^{-\lambda g(t)}>0:\\
(\gamma^{c})^{\prime }& =-\lambda h(t)(1-c)e^{-\lambda g(t)}.
\end{align*}%
Let $\theta (c)$ denote the optimal switching time when the charge is $c.$
Consider two charge levels $c_{L}<c_{H}.$ and corresponding switching times $%
\theta _{L}$ and $\theta _{H}.$ Put $\Delta c:=c_{H}-c_{L}>0$ and compute:%
\begin{align*}
F(c)&:=\int_{0}^{\theta (c)}\gamma^{c}_t\text{ }\mathrm{d}t:\\
F^{\prime }(c)&=\int_{0}^{\theta (c)}\frac{\partial }{\partial c}\gamma
^{c}_t\text{ }\mathrm{d}t+\gamma^{c}(\theta (c))\theta ^{\prime }(c).
\end{align*}%
By the Mean Value Theorem for some $c$ in $(c_{L},c_{H})$
\begin{align*}
&\int_{0}^{\theta _{L}}\gamma^{L}_t\text{ }\mathrm{d}t-\int_{0}^{\theta
_{H}}\gamma^{H}_t\text{ }\mathrm{d}t =F(c_{L})-F(c_{H})\\
&=-[F(c_{L}+\Delta
c)-F(c_{L})]=-\Delta cF^{\prime }(c) \\
& =\int_{0}^{\theta _{H}}(\gamma^{L}_t-\gamma^{H}_t)\text{ }\mathrm{d}%
t+\int_{\theta _{H}}^{\theta _{L}}\gamma^{L}_t\text{ }\mathrm{d}t.
\end{align*}%
So the difference between bond yields corresponding to new disclosures is
\begin{align*}
&\int_{0}^{\theta _{H}}(\gamma^{H}_t-\gamma^{L}_t)\text{ }\mathrm{d}%
t-\int_{\theta _{H}}^{\theta _{L}}\gamma^{L}_t\text{ }\mathrm{d}t\\
&=\Delta c[\gamma
^{L}_t(\theta (c))\theta ^{\prime }(c)+
\int_{0}^{\theta (c)}(1-e^{-\lambda g_{\rho}(t)})\text{ }\mathrm{d}t
].
\end{align*}%
The two bracketted term on the right-hand side of the equation, representing the marginal benefit of voluntary disclosure, may be signed most easily numerically, since the
(negative) value of $\theta ^{\prime }(c)$ is given by a rather complicated
formula. For small $\beta $ this is negative. As $\beta $
increases the expression is initially positive and then negative arising
from $\theta ^{\prime }(c)<0.$ Formally, let $c=\underline{c}$ be the
solution of the equation:
\begin{equation*}
\int_{0}^{\theta (c)}(1-e^{-\lambda g_{\rho}(t)})\text{ }\mathrm{d}t+\gamma
_{c}(\theta (c))\theta ^{\prime }(c)=0.
\end{equation*}%
The claims follow for $c$ in the three ranges $(0,\underline{c}),(\underline{%
c},\bar{c}),(\bar{c},1).$

The proof for the other parameters $v$ and $\rho$ proceeds identically, and likewise requires signing numerically. \hfill $\square $
\section*{Appendix B}

\subsection*{Observations on the call formula}
We begin by recalling the call formula (Bingham and Kiesel, 1998):
\begin{align*}
&\mathbb{E}_{t}^{\mathbb{Q}}[(Y_{1}-\gamma_{1})^{+}|Y_{t} =y]\\
&=y\Phi\left( \frac {%
\log(y/\gamma_{1})+\frac{1}{2}\sigma^{2}(1-t)}{\sigma\sqrt{(1-t)}}\right) \\
& -\gamma_{1}\Phi\left( \frac{\log(y/\gamma_{1})-\frac{1}{2}\sigma^{2}(1-t)}{%
\sigma\sqrt{(1-t)}}\right)
\end{align*}
so $\displaystyle{ \rightarrow y-\gamma_{1}}$  as $y\rightarrow\infty$.
We need only consider $y<\gamma_{1}$ here; then, indeed,%
$$\displaystyle{\left[ 1-\Phi\left( \frac{\log(y/\gamma_{1})+\frac{1}{2}\sigma^{2}(1-t)}{%
\sigma\sqrt{(1-t)}}\right) \right]%
<\left[ 1-\Phi\left( \frac {%
\log(y/\gamma_{1})-\frac{1}{2}\sigma^{2}(1-t)}{\sigma\sqrt{(1-t)}}\right) %
\right]}.$$
So in particular, for $0<y<\gamma_{1},$%
\begin{align*}
&y\left[ 1-\Phi\left( \frac{\log(y/\gamma_{1})+\frac{1}{2}\sigma^{2}(1-t)}{%
\sigma\sqrt{(1-t)}}\right) \right] \\
&<\gamma_{1}\left[ 1-\Phi\left( \frac{%
\log(y/\gamma_{1})-\frac{1}{2}\sigma^{2}(1-t)}{\sigma\sqrt{(1-t)}}\right) %
\right] : \\
y-\gamma_{1}&<\\
&y\Phi\left( \frac{\log(y/\gamma_{1})+\frac{1}{2}\sigma^{2}(1-t)%
}{\sigma\sqrt{(1-t)}}\right)\\
&-\gamma_{1}\Phi\left( \frac{\log (y/\gamma_{1})-%
\frac{1}{2}\sigma^{2}(1-t)}{\sigma\sqrt{(1-t)}}\right) .
\end{align*}
And the inequality continues to hold also for all $y>\gamma_{1},$ though we
do not need this.

\subsection*{Distribution of a drifted Brownian minimum}
The probability of a drifted-Brownian minimum started at $0$ falling below $%
a\leq0$ and the maximum falling below $a\geq0$ are as below (see (Karatzas and Shreve 2014, Section 2.8).
\begin{align*}
&\Pr\left[ \min_{[0,T]}\left( \mu t+W_{t}\right) \leq a\right]\\
&=\Phi\left(
\frac{a-\mu T}{\sqrt{T}}\right) +e^{2\mu a}\Phi\left( \frac{a+\mu T}{\sqrt{T}%
}\right) \\
&\Pr\left[ \max_{[0,T]}\left( \mu t+W_{t}\right) \leq a\right]\\
& =\Phi\left(
\frac{a-\mu T}{\sqrt{T}}\right) -e^{2\mu a}\Phi\left( \frac{-a-\mu T}{\sqrt{T%
}}\right) .
\end{align*}

\section*{Appendix C: Formalities and Mathematica code}

\subsection*{Formal aspects of Equilibrium}
Here we expand the informal equilibrium considerations in Section 3.2 where
we indicated how the firm computes the threshold value for disclosing
observed signals and the representative agent values the firm in silence
according to the Minimum Principle of Section 3.3. The text is based on our
earliest continuous-time development.

Formally, the firm uses its private filtration $\mathbb{Y}^{\text{priv}}$
(time-indexed private information consisting of $\sigma $-algebras generated
by the observations induced by a Poisson process) to selectively
disclose/make public any privately received signals, thereby creating a
public filtration $\mathbb{Y}^{\text{pub}}$. Its principal concern is to
maintain a market valuation of its stock as high as possible knowing the
consequences of non-disclosure. It must therefore at each arrival time of a
signal calculate the market forecast of the mandatory time-$1$ disclosure
using $\mathbb{Y}^{\text{pub}}$ to determine whether to disclose the signal.
Their disclosure strategy must thus select a sub-filtration $\mathbb{G}%
^{\ast }=\{\mathcal{G}_{t}^{\ast }\}$ of $\mathbb{Y}^{\text{pub}}$ based on
a cutoff process $\gamma _{t}$ such that at each time $t\in (0,1)$
\begin{align*}
\mathbb{E}^{\mathbb{Q}}[Y_{1}|\mathcal{G}_{t}^{\ast }]=\sup_{\mathbb{G}}\mathbb{E}^{\mathbb{Q}}[Y_{1}|%
\mathcal{G}_{t}],  \tag{OC}
\end{align*}
the supremum being taken over all sub-filtrations $\mathbb{G}=\{\mathcal{G}%
_{t}\}$ of $\mathbb{Y}^{\text{pub}}.$ (This is something akin to a
fixed-point calculation.) The analysis in (Gietzmann and Ostaszewski, 2016)
finds this optimal censoring problem to reduce to a Conditional Bayes
formula, cf. (Jung and Kwon, 1988). In the single-firm setting, conditional
on there being no disclosure in the interval $(t,s)$, the formula
characterizes the time-$t$ disclosure threshold as $\tilde{\gamma}_{t}:=%
\mathbb{E}_{t}^{\mathbb{Q}}[Y_{1}]$ by the formula
\begin{equation*}
\tilde{\gamma}_{t}=\mathbb{E}_{t}^{\mathbb{Q}}[Y_{1}|ND_{s}(\tilde{\gamma}_{s})],
\end{equation*}%
where $ND_{s}(\gamma )$ is the event of non-disclosure when the cutoff is $%
\gamma $.

We now need a number of formal constructs.

The first is a (c\`{a}dl\`{a}g) Poisson process $N(t)$ with intensity $%
\lambda$, the known private information arrival rate, and is independent of
the signal process $Y_{t}$. The arrival times in $(0,1)$ of the process $N$
are regarded as consecutively numbered, with $\theta_{n}$ denoting the $n$%
-th of these; that is, on setting $\theta_{0}=0$,%
\begin{equation*}
\theta_{n}:=\inf\{t>\theta_{n-1}:N(t)>N(t-)\}.
\end{equation*}

The resulting process $Y^{\text{obs}}$ is the private observation process of
the firm. It is piecewise constant, and defines the private filtration of
the firm. This is the (time-indexed) family $\mathbb{Y}^{\text{priv}}$ of $%
\sigma $-algebras generated by the jumps if $N$ at or before time $t$, and
by the observations at or before time $t$, here regarded as space--time
point processes (i.e. taken together with their dates---for background, see
(Daley et al., 2003) and especially (Daley et al. 2008, Chapter 15); it is
formally given by $\mathbb{Y}^{\text{priv}}:=\{\mathcal{Y}%
_{t}:t\in\lbrack0,1]\}$, where%
\begin{equation*}
\mathcal{Y}_{t}:=\sigma(\{(s,Y_{s},N(s)):0<s\leq t\}).
\end{equation*}

Next, the \textit{public filtrations} $\mathbb{G}^{\text{pub}}$ formalize
the notion of `information disclosed via the publicly observed history of $%
Y_{t}$ '. Consider a fixed marked point process (MPP) comprising the
following list of items.

(i) The underlying c\`{a}dl\`{a}g counting process $N^{\text{pub}}(t)$ (`the
disclosure-time process');

(ii) The functions $\theta_{\pm}^{\text{pub}}$ , where $\theta_{-}^{\text{pub%
}}(t)$ is the last arrival time of $N^{\text{pub}}(t)$ less than or equal to
$t$, and $\theta_{+}^{\text{pub}}(t)$ is the first arrival time of $N^{\text{%
pub}}(t)$ greater than or equal to $t$, relations that hold almost surely on
$\mathcal{F}_{t}$ for every $t$ in $[0,1]$;

(iii) The marks $Y^{\text{pub}}(t)=Y(\theta_{-}^{\text{pub}}(t))$ (`the
corresponding observation at time $t$').

In terms of the given MPP, the publicly observed history, briefly the
\textit{public filtration}, which is right-continuous, is taken to mean $%
\mathbb{G}=\mathbb{G}^{\text{pub}}:=\{\mathcal{G}_{t}^{+}:t\in\lbrack0,1]\}$%
, where $\mathcal{G}_{1}^{+}=\sigma(\mathcal{G}_{1},(1,Y_{1}))$ and, for
each $t$ in $[0,1)$, conventionally,%
\begin{equation*}
\mathcal{G}_{t}^{+}=\bigcap\nolimits_{s>t}\mathcal{G}_{s}
\end{equation*}
with the $\sigma$-algebras $\mathcal{G}_{t}$ generated as the join of the
two $\sigma$-algebras corresponding to items (i), (ii), and (iii), dates
included, namely%
\begin{equation*}
\mathcal{G}_{t}=\bigvee\nolimits_{s\in\lbrack0,t)}\sigma((s,N^{\text{pub}%
}(s)),(s,Y^{\text{pub}}(s)).
\end{equation*}

We further define \textit{disclosure filtrations consistently generated from
the (private) filtration }$\mathbb{Y}^{\text{priv}}$ \textit{via the} $%
\mathbb{G}$\textit{-predictable censor} $\gamma=(\gamma_{t})$, to be public
filtrations satisfying the condition:

(vi) If $N^{\text{pub}}(t)>N^{\text{pub}}(t-),$ then $Y^{\text{pub}%
}(t)=Y_{t} $

(vii) If $N(t)>N(t-)$ and $Y_{t}>\gamma_{t}$, then $N^{\text{pub}}(t)>N^{%
\text{pub}}(t-)$.

For such a filtration $\mathbb{G}^{\text{pub}}$, the counting-process for
arrival times occurring in $(0,1)$ in (ii) will be termed the voluntary
disclosure event times (briefly disclosure times): $\theta_{0}^{\text{pub}}$%
, $\theta_{1}^{\text{pub}}$,... .

Associated with a fixed disclosure filtration $\mathbb{G}$ $=\{\mathcal{G}%
_{t}^{+}\}$, consistently generated via $\gamma$ as above, is the process
obtained by taking contingent expectations of the time-$1$ signal $Y_{1}$
with respect to the time-$t$ information subsets, i.e. $t\mapsto\mathbb{E}^{\mathbb{Q}}%
[Y_{1}|\mathcal{G}_{t}^{+}]$. It is this process that is interpreted as a $\mathbb{G}$-predictable
valuation process of the firm (the $\mathbb{G}$\textit{-forecasting
process}.

The optimisation problem (termed the censoring problem in (Gietzmann and
Ostaszewski, 2016)) calls for the construction of a
filtration $\mathbb{G}^{\ast}=\{\mathcal{G}_{t}^{\ast}\}$, necessarily
unique, whose associated forecasting process is the left-sided-in-time
pointwise supremum over all $\mathbb{G}$-forecasting processes, that is, for
each time $t\in(0,1)$, equation (OC) above holds.

In (OC) both sides of the equation
depend only on the public information available to the left of the date $t$.
By definition, such a $\mathbb{G}^{\ast}$, if it exists, is unique. The main
results in (Gietzmann and Ostaszewski 2016, Theorem 1), assert in particular
that the unique censor process $\gamma_{t}$ corresponding to the optimal $%
\mathbb{G}$\textit{-}forecasting process has the form $\sup _{\mathbb{G}}%
\mathbb{E}^{\mathbb{Q}}[Y_{1}|\mathcal{G}_{t}]$ and satisfies the (censoring)
differential equation%
\begin{equation*}
\gamma_{t}^{\prime}=-\gamma_{t}\nu_{t}\text{ for }\theta_{n-1}^{\text{pub}%
}<t<\theta_{n}^{\text{pub}},
\end{equation*}
where $\nu_{t}=\lambda\lbrack2\Phi(\frac{1}{2}\sigma\sqrt{1-t})-1]>0.$
\bigskip

\textbf{Remarks} 1. Informally, the censoring problem requires the firm to
reach disclosure/suppression decisions using only the history of all prior
public information.

2. The optimal censoring problem amounts to the construction of a censoring
filter (process), which for short we call a censor, $\gamma_{t}$ with the
following properties:
\medskip

\textbullet\ the disclosure subfiltration $\mathbb{G}$ of $\mathbb{Y}^{\text{%
priv}}$, generated through suppression of observations by reference to the
process $\gamma_{t}$, is consistently generated from the private filtration $%
\mathbb{Y}^{\text{priv}}$ via $\gamma_{t}$;
\medskip

\textbullet\ $\gamma$ is $\mathbb{G}$-predictable (this is crucial);
\medskip

\textbullet\ for each time instant $t\in(0,1)$, $\gamma_{t}$ is chosen to be
a cutoff maximizing the expected value of $Y_{1}$, given only past public
information (which is why $\gamma$ needs to be $\mathbb{G}$-predictable).

3. The role of $\mathbb{G}$ above is to formalize the censoring of the
private observation process relative to information `public' before any time-%
$t$ disclosure, so in general distinct from the optional valuation process $%
\mathbb{E}^{\mathbb{Q}}[Y_{1}|\mathcal{G}_{t}^{+}]$, which models the later
right-continuous public valuation at time $t.$

4. In Remark 2 the agent is a maximizer of an instantaneous objective linked
to the terminal signal $Y_{1}$ (the legally defined accounting estimation of
firm value, reported at the mandatory date $1$). The alternative approach is
to establish a single overall performance indicator for the entire
trajectory of the estimator. Optimality of overall economic behaviour
induced by instantaneous (sometimes called `myopic') objectives is
established for a class of models related to ours in Feldman (1992). Compare
also Bj\"{o}rk and Murgoci (2014).

\subsection*{Mathematica Model code}

Phi[x\_] := 0.5(1 + Erf[x/Sqrt[2]])

\noindent phi[x\_] := Exp[-0.5x\symbol{94}2]/Sqrt[2*Pi]

\noindent h[t\_, a\_] := 2*Phi[a*Sqrt[1 - t]] - 1

\noindent \text{ }\% a=s/2

\noindent asq[t\_,a\_]:=a*Sqrt[1-t]

\noindent H[t\_,a\_]:=Phi[a]+(-Phi[a]+a*phi[a])/a\symbol{94}2-((asq[t,a]%
\symbol{94}2-1)*Phi[asq[t,a]]+asq[t,a]*phi[asq[t,a]])/a\symbol{94}2 -t/2;

\noindent
candTheta[s\_,L\_,kappa\_]:=FindRoot[H[t,s/2]==(-Log[kappa]/L),\{t, 0.4,
0.5\}][[1, 2]];

\noindent sparTheta[s\_, L\_, kappa\_] := FindRoot[(1 - t)*h[t,s/2] == (kappa%
\symbol{94}\{-1\} - 1)/(L), \{t, 0.4, 0.5\}][[1, 2]]

\noindent sparLowest[s\_,kappa\_]=((kappa\symbol{94}\{-1\} -
1)/h[0,s/2])[[1]],

\noindent candLowest[s\_,kappa\_]=(-Log[kappa])/H[1,s/2]),

\subsection*{Stats code}
\textit{Implementation tables below assume }$s=\sigma =3$\textit{\ and }$%
\kappa =0.7$\textit{.}

\noindent $<<$ Statistics`ContinuousDistributions'

\noindent $<<$ Statistics`DiscreteDistributions'

\noindent $<<$ NumericalMath`SplineFit'

\noindent minbm[m\_,s\_,L\_,t\_]:=Module[\{nn,xx,xx0,xx1, yy,yy1\},

\noindent nn=Random[PoissonDistribution[L]];

\noindent If[nn==0,xx=Table[\{0\}],xx =
Sort[RandomArray[UniformDistribution[0, 1], nn]];

\noindent xx0=Prepend[xx,0];

\noindent xx1=Table[xx0[[k]]-xx0[[k-1]],\{k,2,Length[xx0]\}];

\noindent If[nn==0,yy=Table[\{0\}],yy = RandomArray[NormalDistribution[0,
1], nn]];

\noindent yy1=yy*s*Sqrt[xx1]+1*xx1;

\noindent yy2=Table[Sum[yy1[[j]],\{j,1,k\}],\{k,1,Length[yy1]\}];

\noindent Min[yy2]]

\bigskip
\noindent candTarget[s\_, L\_, kappa\_] := 0.5s*candTheta[s,L, kappa] -
L(H[1, s/2] - H[candTheta[s,L, kappa], s/2])/(s/2)

\noindent candidPr[m\_, s\_, t\_, L\_, kappa\_, nn\_] := Module[\{dt, dmi\},

\noindent dt = Sort[Table[minbm[m, s, L, t], \{k, 1, nn\}]]; dmi = Min[dt];

\noindent RangeCounts[dt, \{dmi, candTarget[s, L, kappa]\}][[2]]/Length[dt]]

\bigskip
\noindent candPts = Table[\{L, (1 - Exp[-L(1 -
candTheta[L,0.7])])*candidPr[-0.5*3, 3, candTheta[s,L, 0.7], L, 0.7, 200]\},
\{L,2, 11\}];

\noindent splineCand=SplineFit[candPts,Bezier];

\noindent ParametricPlot[splineCand[u],\{u,0,9\}]

\noindent sparTarget[s\_, t\_, L\_] := -2(L/s)*H[t, s/2]

\noindent sparingPr[m\_, s\_, t\_, L\_, kappa\_, nn\_] := Module[\{dt, dmi\},

\noindent dt = Sort[Table[minbm[m, s, L, t], \{k, 1, nn\}]]; dmi = Min[dt];

\noindent RangeCounts[dt, \{dmi, sparTarget[s, 1, L]\}][[2]]/Length[dt]]

\bigskip
\noindent L[i\_] := i/10;

\noindent sparPts = Table[\{i/10, (1 - Exp[-L[i]*sparTheta[3, L[i],
0.7]])*sparingPr[-0.5*3, 3, sparTheta[3, L[i], 0.7], L[i], 0.7, 2000]\},
\{i, 1, 121\}];

\noindent splineSpar=SplineFit[sparPts,Bezier];

\noindent ParametricPlot[splineSpar[u],\{u,1,100\}]

\end{document}